\documentclass[10pt]{revtex4}
\usepackage{amssymb}
\usepackage{latexsym}
\usepackage{epsfig}
\usepackage{float}
\usepackage{graphicx,epsfig, color }
\usepackage{graphicx}
\usepackage{subfigure}
\usepackage{amsmath}
\begin{document}
\title{Reissner-Nordstr\"om Black Holes in Quartic Quasi-Topological Gravity Theory}
\author{ M. Ghanaatian$^{1}$\footnote{Corresponding author}, F. Naeimipour$^{1}$, A. Bazrafshan$^{2}$, M. Abkar$^{1}$,}
\address{$^1$ Department of Physics, Payame Noor University (PNU), P.O. Box 19395-3697 Tehran, Iran\\
$^2$ Department of Physics, Jahrom University, 74137-66171 Jahrom, Iran}

\begin{abstract}
In this paper, we construct the exact solutions of Reissner-Nordstr\"om black holes in the presence of quartic quasi-topological gravity. we obtain the thermodynamics and conserved quantities of the solutions and check the first law of thermodynamics. In studying the physical properties of the solutions, we consider asymptotically Ads, dS and flat solutions of Reissner-Nordstr\"om black hole in quartic quasi-topological gravity and compare them with Einstein and third-order quasi-topological gravities. we also investigate the thermal stability of the solutions that we show the thermal stability are just for AdS solutions not for dS and flat ones.

\end{abstract}



\maketitle

\section{Introduction}
There are several motivations to study modified gravity with higher curvature terms. One of them is related to AdS/CFT which argues that a certain conformal field theory in (d+1) dimensions $(CFT_{d+1})$ corresponds to the (super)gravity on (d+2) dimensional anti de-Sitter space $(AdS_{d+2})$ \cite{Malda1,Malda2}. This correspondence is a miracle to solve many problems in an easy way which solving them in CFT is so hard or unresolvable, like entanglement entropy in high dimensions \cite{entang}. Hilbert Einstein is the most simple gravity which is only dual to those conformal field theories for which all the central charges are equal. However it does not have enough free parameters to relate them to the central charges of CFT \cite{Bazr}.

Also, the results of some holographic constructions can show that the ratio of the shear viscosity to entropy
density is not in accordance with what is obtained in CFT's correspondence to Einstein gravity \cite{Shear1,Kov,Shear2,Lands,Mateos}. However, perturbation of Einstein gravity can lead to a class of CFT’s
in which this ratio generally depends on the value of the additional gravitational couplings \cite{Mye}. These reasons attract people to go to the modified theories such as Lovelock \cite{Friedman,Schleich,Jacobs} or quasi-topological theory \cite{Cai1,Mann1,Lemos1,Lemos2,Lemos3,Brenna,Dehghani1,Dehghani2,Dehghani3,Aminn}. In spherically symmetry conditions, these two theories are almost similar. For example, the obtained solutions of both theories should get to the Einstein's solutions in the absence of new couplings, but because the quasi-topological terms are not true topological invariants, so this gravity has the ability to produce effective gravitational effects in fewer dimensions than Lovelock gravity. For example, while the cubic Lovelock gravity acts in seven and higher dimensions, the quasi-topological one acts effectively in five and higher ones. Also, as the CFT should obey causality, this causes constraint on the coupling constant. Unlike Lovelock theory, the coupling of cubic quasi-topological are defined in a way that causality happens \cite{Myer1,Brigante,Sin,Ge,Camanho,Hofman}.

On maximally symmetric backgrounds, quasi-topological gravity can cause linearized equations of motion coincide with
the linearized Einstein equations. This causes physical meanings in two sides. First, on the vacuum, some of extra degrees of freedom over the ones in Einstein gravity are ghost which have negative kinetic energy and they means a breakdown of unitarity in the quantum theory \cite{Sisman}. Second, holographic studies of the theory are too simplified because on these backgrounds, the theory propagates the same degrees of freedom as Einstein’s gravity \cite{Paulos}. Recently, holographic p-wave superconductor has been studied in the quasi-topological gravity in the probe limit. The obtained data of this theory is suitable for the Drug model in the low-frequency limit \cite{Kuang}.

Recently, the idea of an action quartic in curvature terms has been done in \cite{Oliva}, but it was not successful because the field equations in dimensions less than seven were vanished. This was a motivation to construct the idea of quartic quasi-topological which can be explained in all dimensions more than four except 8. In this theory, by adding a new coupling constant (there are four coupling constants), the constraints appearing from causality may not identify the three constraints appearing from necessary positive energy fluxes.

A lot of studies in quartic quasi-topological have been done\cite{Bazr,Bazr1,Ghanna}. For example, quartic quasi-topological gravity in the spherically symmetric case has been studied in \cite{Bazr}. Some of the effects of quartic quasi-topological term for Lifshitz-symmetric black holes have been investigated \cite{Ghanaa1}. A review of quartic quasi-topological black holes in the presence of a nonlinear electromagnetic Born-Infeld field is presented in \cite{Ghanaa2}.
Now we are pleased to study Reissner-Nordstr\"om black holes in quartic quasi-topological gravity.
This paper is arranged as this: we first obtain the field equations of these black holes. Then we find the exact solutions of these equations. In section \ref{thermo}, we obtain thermodynamics and conserved quantities of the solutions and check the first law of thermodynamics. Then, in section \ref{physical}, we investigate the physical properties and structure of the solutions. At last, we have a brief study on the whole paper and obtained results.

\section{Field equations and solutions}\label{Field}
We start with the (n+1)-dimensional action in quasi topological gravity in the presence of fourth order curvature correction
\begin{equation}\label{Act1}
S=\frac{1}{16\pi}\int{d^{n+1}x\sqrt{-g}\big\{-2\Lambda+{\mathcal L}_1+\frac{\lambda L^2}{(n-2)(n-3)}{\mathcal L}_2+\frac{7\mu L^4}{4}{\mathcal L}_3+\frac{cL^6}{21024}{\mathcal L}_4-\frac{1}{4}F^2\big\}},
\end{equation}
where $g$ is the determinant of metric ($g_{\mu\nu}$) and  $\Lambda$ is the cosmological constant. We define $F^2=F_{\mu\nu}F^{\mu\nu}$ where $F_{\mu\nu}$ is the electromagnetic field tensor that is defined as   $F_{\mu\nu}=\partial_{\mu}A^{\nu}-\partial_{\nu}A^{\mu}$ and $A_{\mu}$ is the vector potential. \\
${\mathcal L}_1=R$ is the Einstein-Hilbert Lagrangian and ${\mathcal L}_2=R_{abcd}R^{abcd}-4R_{ab}R^{ab}+R^2$ is the second order Lovelock (Gauss-Bonnet) Lagrangian. The third and fourth order correction in quasi-topological gravity are:
\begin{eqnarray}
{{\mathcal L}_3}&=&
R_a{{}^c{{}_b{{}^d}}}R_c{{}^e{{}_d{{}^f}}}R_e{{}^a{{}_f{{}^b}}}+\frac{1}{(2n-1)(n-3)} \bigg(\frac{3(3n-5)}{8}R_{abcd}R^{abcd}R-3(n-1)R_{abcd}R^{abc}{{}_e}R^{de}\nonumber\\
&&+3(n+1)R_{abcd}R^{ac}R^{bd}+6(n-1)R_a{{}^b}R_b{{}^c}R_{c}{{}^a}-\frac{3(3n-1)}{2}R_a{{}^b}R_b{{}^a}R +\frac{3(n+1)}{8}R^3\bigg),
\end{eqnarray}

\begin{eqnarray} {\mathcal{L}_4}&=& c_{1}R_{abcd}R^{cdef}R^{hg}{{}_{ef}}R_{hg}{{}^{ab}}+c_{2}R_{abcd}R^{abcd}R_{ef}{{}^{ef}}+c_{3}RR_{ab}R^{ac}R_c{{}^b}+c_{4}(R_{abcd}R^{abcd})^2\nonumber\\
&&+c_{5}R_{ab}R^{ac}R_{cd}R^{db}+c_{6}RR_{abcd}R^{ac}R^{db}+c_{7}R_{abcd}R^{ac}R^{be}R^d{{}_e}+c_{8}R_{abcd}R^{acef}R^b{{}_e}R^d{{}_f}\nonumber\\
&&+c_{9}R_{abcd}R^{ac}R_{ef}R^{bedf}+c_{10}R^4+c_{11}R^2 R_{abcd}R^{abcd}+c_{12}R^2 R_{ab}R^{ab}\nonumber\\
&&+c_{13}R_{abcd}R^{abef}R_{ef}{{}^c{{}_g}}R^{dg}+c_{14}R_{abcd}R^{aecf}R_{gehf}R^{gbhd},
\end{eqnarray}
with the definition,
\begin{eqnarray}
&&c_{1}=-(n-1)(n^7-3n^6-29n^5+170n^4-349n^3+348n^2-180n+36)\nonumber\\
&&c_{2}=-4(n-3)(2n^6-20n^5+65n^4-81n^3+13n^2+45n-18)\nonumber\\
&&c_{3}=-64(n-1)(3n^2-8n+3)(n^2-3n+3)\nonumber\\
&&c_{4}=-(n^8-6n^7+12n^6-22n^5+114n^4-345n^3+468n^2-270n+54)\nonumber\\
&&c_{5}=16(n-1)(10n^4-51n^3+93n^2-72n+18)\nonumber\\
&&c_{6}=-32(n-1)^2(n-3)^2(3n^2-8n+3)\nonumber\\
&&c_{7}=64(n-2)(n-1)^2(4n^3-18n^2+27n-9)\nonumber\\
&&c_{8}=-96(n-1)(n-2)(2n^4-7n^3+4n^2+6n-3)\nonumber\\
&&c_{9}=16(n-1)^3(2n^4-26n^3+93n^2-117n+36)\nonumber\\
&&c_{10}=n^5-31n^4+168n^3-360n^2+330n-90\nonumber\\
&&c_{11}=2(6n^6-67n^5+311n^4-742n^3+936n^2-576n+126)\nonumber\\
&&c_{12}=8(7n^5-47n^4+121n^3-141n^2+63n-9)\nonumber\\
&&c_{13}=16n(n-1)(n-2)(n-3)(3n^2-8n+3)\nonumber\\
&&c_{14}=8(n-1)(n^7-4n^6-15n^5+122n^4-287n^3+297n^2-126n+18).\nonumber\\
\end{eqnarray}

We would like to find the solutions by this metric
\begin{eqnarray}\label{metr}
ds^2=-\frac{r^2}{L^2}f(r)dt^2+\frac{L^2}{r^2 g(r)}dr^2+\frac{r^2}{L^2} d\Omega^2,
\end{eqnarray}
where $L$ is a scale factor related to the cosmological constant and $f(r)$ and $g(r)$ are the metric functions that should be found. $d\Omega^2$ shows the line element of an $(n-1)$-dimensional hypersurface with constant curvature $(n-1)(n-1)k$ with the volume $V_{n-1}$
\begin{equation}
d\Omega^2=\left\{
\begin{array}{ll}
$$d\theta^{2}_{1}+\sum_{i=2}^{n-1}\prod_{j=1}^{i-1} sin^2 \theta_{j} d\theta_{i}^2$$,\quad \quad\quad\quad \quad\quad\quad\quad\quad\quad  \ {k=1,}\quad &  \\ \\
$$\sum_{i=1}^{n-1} d\phi_{i}^{2}$$,\quad\quad\quad\quad\quad\quad \quad\quad\quad\quad\quad\quad\quad\quad\quad\quad\quad\quad  \ {k=0,}\quad &  \\ \\
$$d\theta^{2}_{1}+sinh^2 \theta_1 d\theta_{2}^{2}+sinh^2 \theta_{1} \sum_{i=3}^{n-1}\prod_{j=2}^{i-1} sin^2 \theta_{j} d\theta_{i}^2$$, \quad{k=-1.}\quad &
\end{array}
\right.
\end{equation}
where the parameter $k = �-1, 0, 1$ corresponding to hyperbolic,
flat and spherical geometries, respectively. To have static solutions, we define vector potential as
\begin{eqnarray}\label{h1}
A_{\mu}=q\frac{r}{L}h(r)\delta_{\mu}^{0},
\end{eqnarray}
where $h(r)$ tends to unity at $r\rightarrow\infty$.
Evaluating the action (\ref{Act1}) with metric (\ref{metr}) and integrating by part leads to the following action
\begin{eqnarray}\label{Act2}
S=\frac{(n-1)}{16\pi L^2}\int d^{n} x\int{dr \sqrt{\frac{f}{g}}\bigg\{\bigg[r^n\bigg(-\frac{\Lambda}{n(n-1)}L^2-\Psi+\lambda\Psi^2+\mu\Psi^3+c\Psi^4\bigg)\bigg]^{'}+\frac{q^2r^{n-1}g}{2(n-1)f}(rh^{'}+h)^2\bigg\}},
\end{eqnarray}
where $\Psi=\bigg(g-\frac{L^2}{r^2}k\bigg)$ and a prime $(')$ shows the derivative with respect to the radial coordinate $r$. We will examine a 5-dimensional gravity theory by substituting $f(r)=N^2(r)g(r)$ and then varying the action (\ref{Act2}) with respect to $f(r)$, $g(r)$ and $h(r)$. They yield respectively to the equations
\begin{equation}\label{equ1}
(-1+2\lambda \Psi+3\mu \Psi^2+4c\Psi^3)N^{'}=0
\end{equation}
\begin{equation}\label{equ2}
\bigg\{3r^4\bigg(-\frac{\Lambda}{6}L^2-\Psi+\lambda \Psi^2+\mu \Psi^3+c\Psi^4\bigg)\bigg\}^{'}=\frac{q^2r^3}{2}\bigg\{\bigg(\frac{(rh)^{'}}{N}\bigg)^2\bigg\}
\end{equation}
\begin{equation}\label{equ3}
\bigg(\frac{r^3}{N}(rh)^{'}\bigg)^{'}=0
\end{equation}
To find the solutions, we start with equation (\ref{equ1}). This shows that N(r) should be constant, So we choose $N(r)=1$. Using this constraint in (\ref{equ3}) and solving this equation causes
\begin{eqnarray}\label{h2}
h(r)=\frac{Q}{2\sqrt{2}L}\frac{1}{r^2},
\end{eqnarray}
where $Q$ is related to the electric charge of the black hole obtaining by using the Gauss law as
\begin{eqnarray}\label{charge}
q=\frac{Q}{16\sqrt{2}\pi L}.
\end{eqnarray}
Using $N=1$ and (\ref{h2}) in (\ref{equ2}) leads to
\begin{eqnarray}\label{equasli}
c \Psi^4+\mu \Psi^3+\lambda \Psi^2-\Psi+\kappa=0
\end{eqnarray}
that $\kappa$ is
\begin{eqnarray}
\kappa=-\frac{\Lambda}{6}L^2-\frac{M}{3r^4}+\frac{Q^2}{24r^6},
\end{eqnarray}
which $M$ is a constant of integration relating to the mass of black hole. This geometrical mass of black hole is
\begin{eqnarray}
M=3r_{+}^4\bigg(-\frac{\Lambda}{6}L^2+\frac{Q^2}{24r_{+}^6}+k\frac{L^2}{r_{+}^2}+\lambda k^2\frac{L^4}{r_{+}^4}-\mu k^3\frac{L^6}{r_{+}^6}+c k^4\frac{L^8}{r_{+}^8}\bigg),
\end{eqnarray}
where $r_{+}$ is defined as the radial coordinate of the outermost horizon of the black hole which is the positive root of $g(r_{+})=0$.
In order to have the real solutions of quartic quasi-topological Rissner-Nordstrom, the following inequality is required \cite{BMRN}
\begin{eqnarray}\label{Delta}
\Delta=\frac{H^2}{4}+\frac{P^3}{27}>0,
\end{eqnarray}
which $P$ and $H$ are
\begin{eqnarray}
\alpha&=&\frac{-3\mu^2}{8c^2}+\frac{\lambda}{c}\,\,\,\,\,\,\,\,\,\,\,,\,\,\,\,\,\,
\beta=\frac{\mu^3}{8c^3}-\frac{\mu\lambda}{2c^2}-\frac{1}{c}\nonumber\\
&&\gamma=\frac{-3\mu^4}{256c^4}+\frac{\lambda \mu^2}{16c^3}+\frac{\mu}{4c^2}+\frac{\kappa}{c},
\end{eqnarray}
\begin{eqnarray}
P&=&-\frac{\alpha^2}{12}-\gamma\,\,\,\,\,\,\,\,\,\,,\,\,\,\,\,\,\,
H=-\frac{\alpha^3}{108}+\frac{\alpha \gamma}{3}-\frac{\beta^2}{8}.
\end{eqnarray}
If we use the following definitions,
\begin{eqnarray}
U=\bigg(-\frac{H}{2}\pm\sqrt{\Delta}\bigg)^{\frac{1}{3}},
\end{eqnarray}
\begin{equation}
y=\left\{
\begin{array}{ll}
$$-\frac{5}{6}\alpha+U-\frac{P}{3U}$$,\quad \quad\quad\quad \  \ {U\neq 0,}\quad &  \\ \\
$$-\frac{5}{6}\alpha+U-\sqrt[3]{H}$$, \quad \quad\quad\quad{U=0,}\quad &
\end{array}
\right.
\end{equation}
\begin{eqnarray}
W=\sqrt{\alpha+2y},
\end{eqnarray}
then the solutions are obtained as
\begin{eqnarray}
f(r)=\frac{L^2}{r^2}k-\frac{\mu}{4c}+\frac{\pm_{s}W\mp_{t}\sqrt{-(3\alpha+2y\pm_{s}\frac{2\beta}{W})}}{2},
\end{eqnarray}
Where the two $\pm_{s}$ should both have the same sign, while the sign of $\pm_{t}$ is independent.

\section{Thermodynamics of the solutions} \label{thermo}

The effort to understand the statistical mechanics of black holes has had a deep impact upon the understanding of quantum gravity and leading to the formulation of the holographic principle. In this part, we study about the available thermodynamic quantities at event horizons of the black hole in order to investigate their stability. Starting from Bekenstein-Hawking entropy theorem, conjectured that the black hole entropy is proportional to the area of its event horizon \cite{Iyer}
\begin{eqnarray}
S=\frac{r_{+}^3}{4}\bigg(1+6\lambda k \frac{L^2}{r_{+}^{2}}+9\mu k^2 \frac{L^4}{r_{+}^{4}}-4ck^3\frac{L^6}{r_{+}^6}\bigg).
\end{eqnarray}
We can obtain temperature by analytic continuation of the metric.
In this method, we use $t\rightarrow i\tau$ for
the Euclidean section of the metric. For regularity at $r = r_{+}$, we should identify
$\tau\rightarrow \tau+\beta_{+}$, where $\beta_{+}$ is the inverse Hawking temperature. This temperature is obtained at the horizon $r_{+}$ as
\begin{eqnarray}\label{temperaure}
T_{+}=\bigg(\frac{r^2 g^{'}}{4\pi L^2}\bigg)_{r=r_{+}}=\frac{1}{6\pi L^2}\frac{64 \pi^2 L^2 q^2 r_{+}^2-6 \mu_{0} r_{+}^8+6ck^4L^8-3\mu k^3 L^6 r_{+}^2-3kL^2r_{+}^6 }{-4c k^3 L^6 r_{+}+3\mu k^2 L^4 r_{+}^3-2\lambda k L^2 r_{+}^5-r_{+}^7}.
\end{eqnarray}
The electric potential $U$, measured at infinity with respect
to the horizon is defined by
\begin{eqnarray}
U=A_{\mu }\chi ^{\mu }\left| _{r\rightarrow \infty }-A_{\mu }\chi
^{\mu }\right| _{r=r_{+}},  \label{Pot}
\end{eqnarray}
where $\chi=\partial_{t}$ is the null generator of the horizon. Using the bove, we
can obtain electric potential
\begin{eqnarray}\label{potential}
U=\frac{Q}{2\sqrt{2} L r_{+}^2}
\end{eqnarray}
There are several ways for calculating the mass of the black hole. One of them is subtraction method in which, we write our metric in the form of
\begin{eqnarray}\label{ADM1}
ds^2=-W^2(r)dt^2+\frac{dr^2}{V^2(r)}+r^2d\Omega^2,
\end{eqnarray}
then the quasilocal mass is obtained as
\begin{eqnarray}
\mathcal M=r W(r)(V_{0}(r)-V(r)).
\end{eqnarray}
It is mentioned that $V_{0}(r)$ is zero of the energy that depends on the choice of reference background. The ADM mass is obtained when $r\rightarrow\infty$ in $\mathcal M(r)$. Changing metric (\ref{metr}) in
the form (\ref{ADM1}), the mass is obtained as
\begin{eqnarray}
m=\frac{M}{16 \pi L^2}
\end{eqnarray}
To check the first law of thermodynamics, if we consider $S$ and $Q$ as a complete set of
extensive parameters for the mass $m(S,q)$, then their intensive parameters are respectively defined as
\begin{eqnarray}\label{firstlaw1}
T=\bigg(\frac{\partial m}{\partial S}\bigg)_{q}\,\,\,\,\,,\,\,\,\,\,U=\bigg(\frac{\partial m}{\partial q}\bigg)_{S}.
\end{eqnarray}
If we calculate the above intensive quantities, they are coincided with Eqs. \eqref{temperaure} and \eqref{potential}. This result shows that these quantities satisfy the first
law of black hole thermodynamics,
\begin{eqnarray}
dM=TdS+UdQ.
\end{eqnarray}

\section{physical Properties of the solutions } \label{physical}
To have a better understanding of $f(r)$, we have plotted this function versus $r$ in this part to investigate them. In all figures (\ref{fig1})-(\ref{fig10}), we have obeyed the condition (\ref{Delta}) to have real solutions. We have also considered $L=1$ for easy. we should mention that for simplicity, we have abbreviated Fourth order Quasi Topological to FQT and Third order Quasi Topological to TQT.
We will see that for $r\rightarrow 0$, All the solutions $f(r)$ in FQT theory have the same behavior and they go to $\infty$.

If $\Lambda$ is negative, positive or 0, then the obtained solutions are respectively asymptotically Anti de Sitter (AdS), de Sitter (dS) or flat. So we have separated the study on $f(r)$ in three bellowing subsection: Asymptotically Anti-de Sitter spacetimes, Asymptotically de Sitter spacetimes and Asymptotically flat spacetimes.

\subsection*{Asymptotically Anti-de Sitter spacetimes}

\begin{figure}
\center
\includegraphics[scale=0.5]{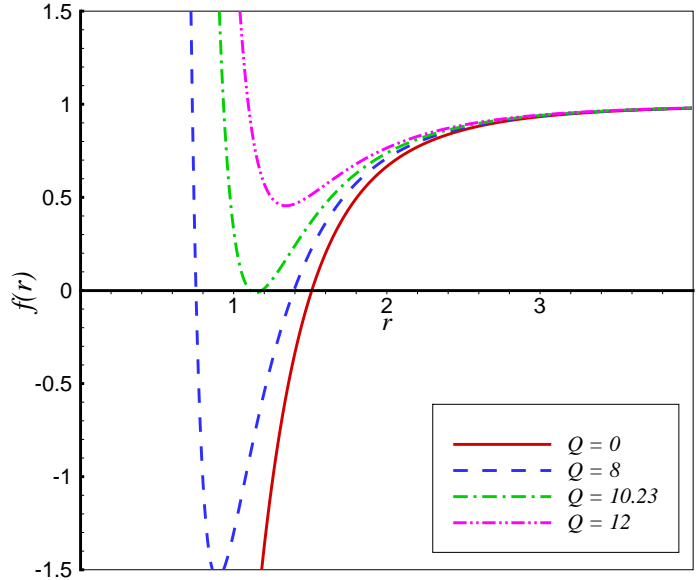}

\caption{\small{Asymptotically AdS solutions of Reissner-Nordstr\"om black hole in FQT gravity with $M=15$, $k=0$, $\lambda=0.04$, $\mu=-0.001$ and $c=-0.0002$.} \label{fig1}}
\end{figure}

\begin{figure}
\center
\includegraphics[scale=0.5]{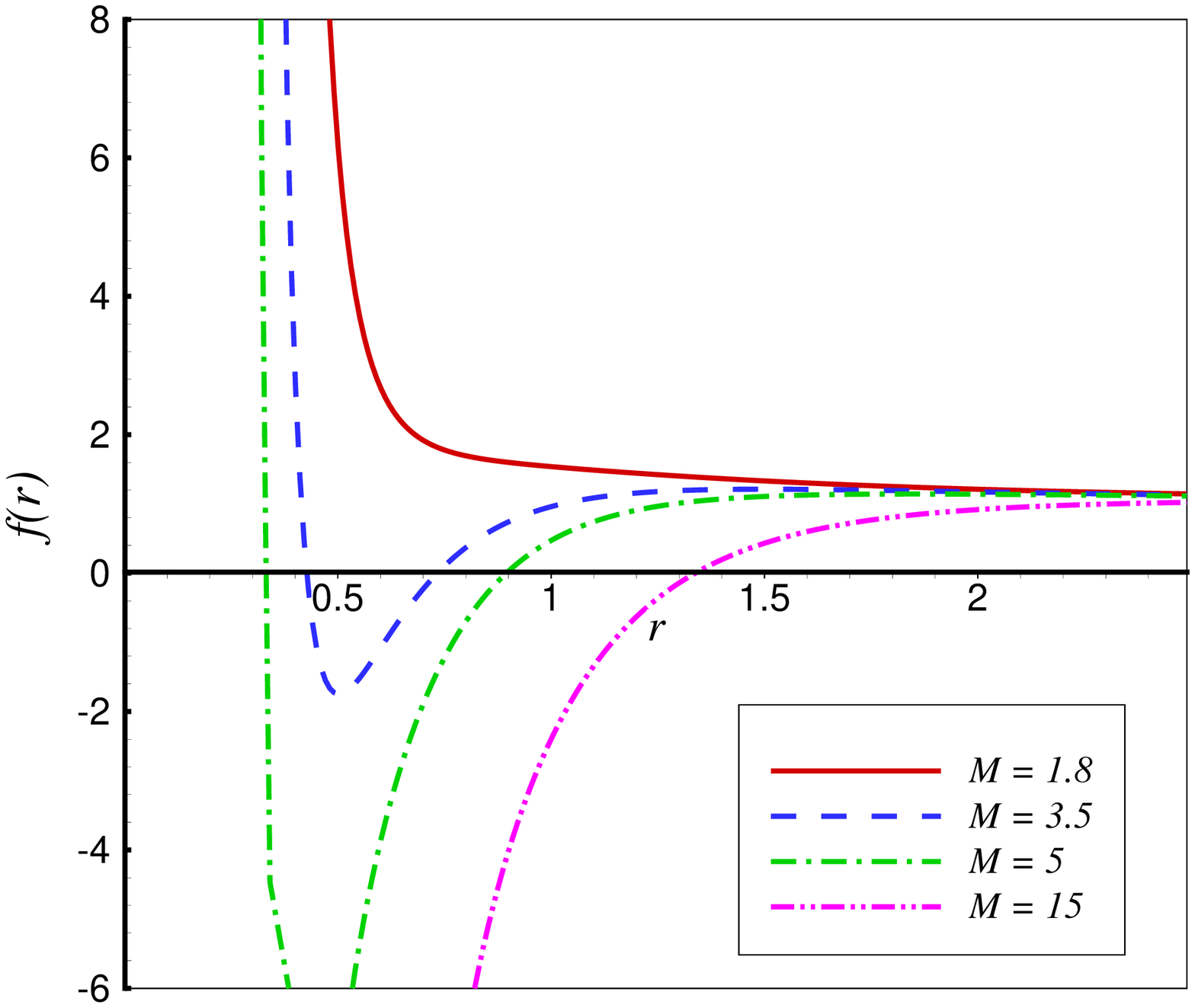}

\caption{\small{Asymptotically AdS solutions of Reissner-Nordstr\"om black hole in FQT gravity with $Q=2$, $k=1$, $\lambda=0.04$, $\mu=-0.001$ and $c=-0.0002$.} \label{fig2}}
\end{figure}

\begin{figure}
\center
\includegraphics[scale=0.5]{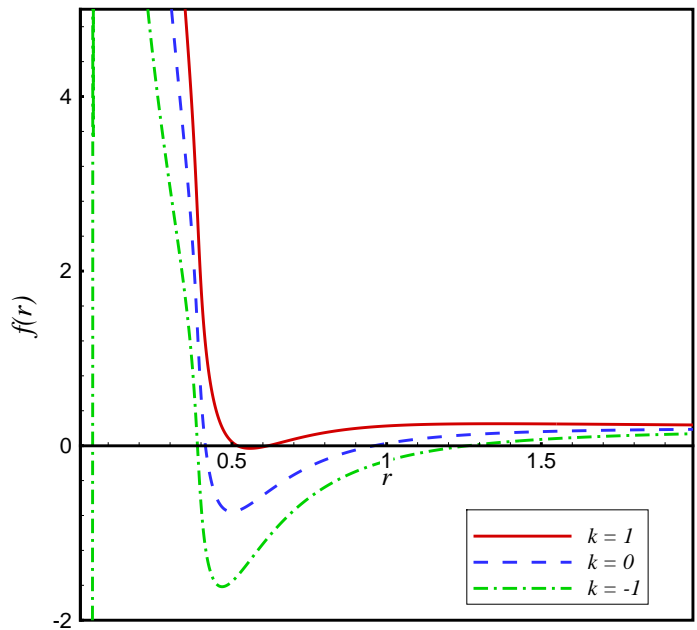}
\caption{\small{Asymptotically AdS solution $f(r)/5$ versus $r$ in FQT gravity with $M=3$, $Q=2$, $\lambda=0.04$, $\mu=-0.001$ and $c=-0.0002$.} \label{fig3}}
\end{figure}

\begin{figure}
\center
\includegraphics[scale=0.5]{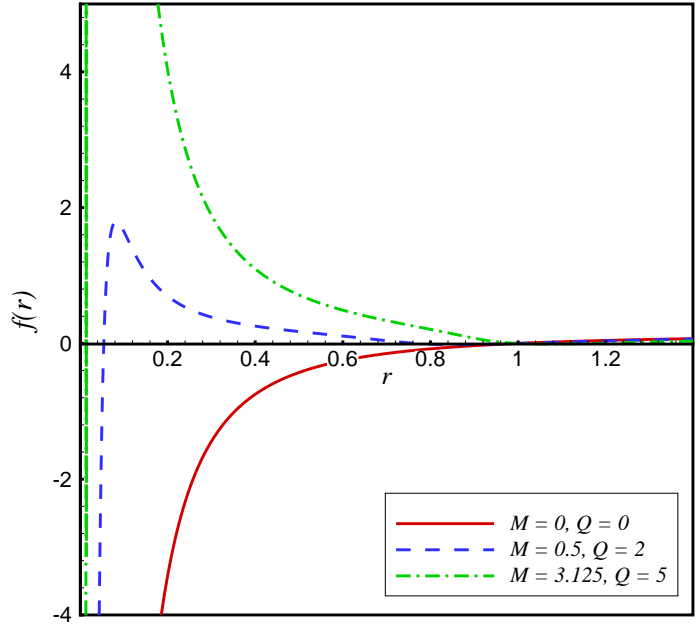}
\caption{\small{Asymptotically AdS solution $f(r)/7$ versus $r$ in FQT gravity with $r_{+}=1$, $k=-1$, $\lambda=0.4$, $\mu=-0.1$ and $c=-0.0002$} \label{fig4}}
\end{figure}

\begin{figure}
\centering
\subfigure[Einstein Gravity]{\includegraphics[scale=0.27]{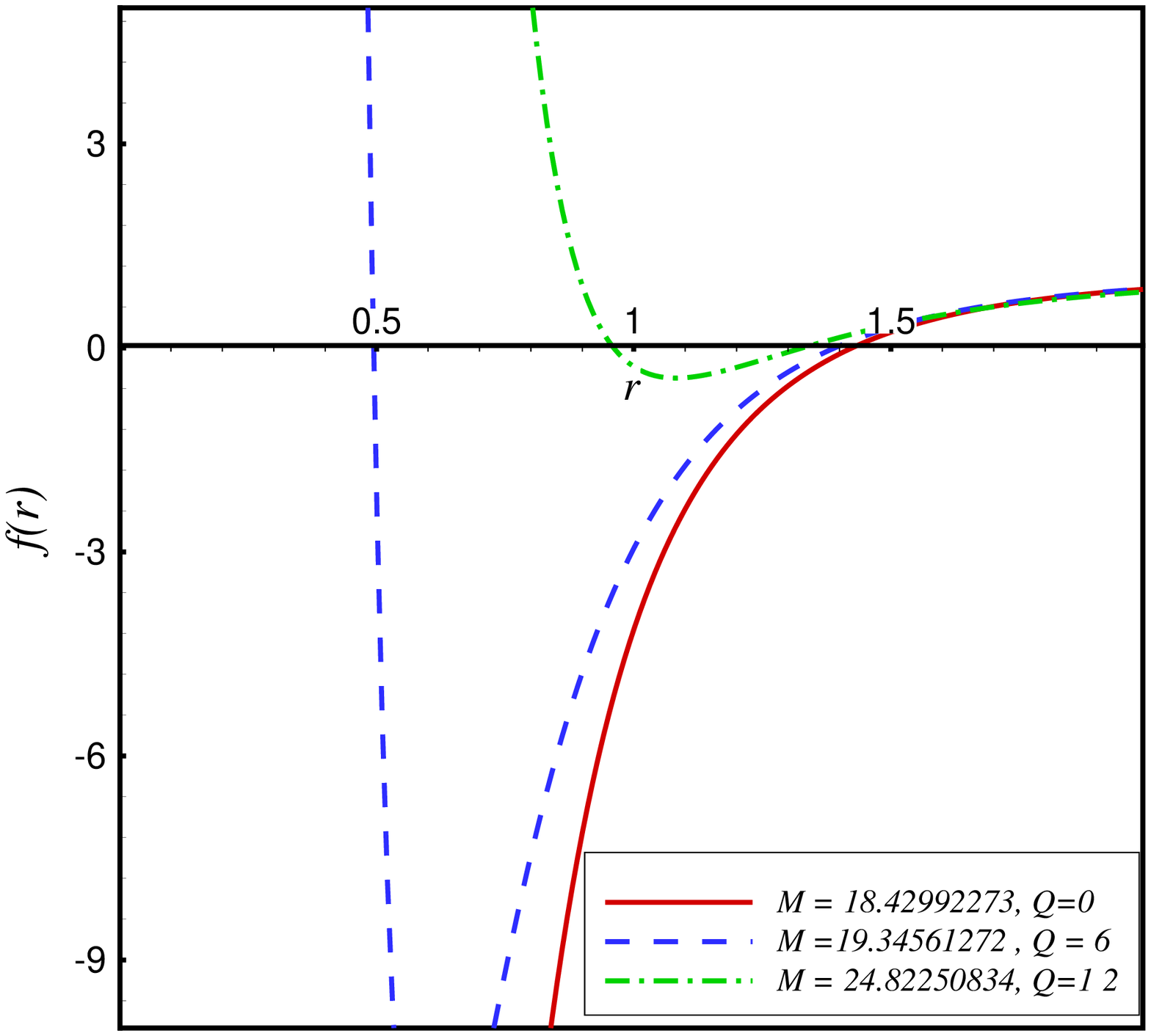}\label{fig5a}}\hspace*{.2cm}
\subfigure[FQT Gravity]{\includegraphics[scale=0.27]{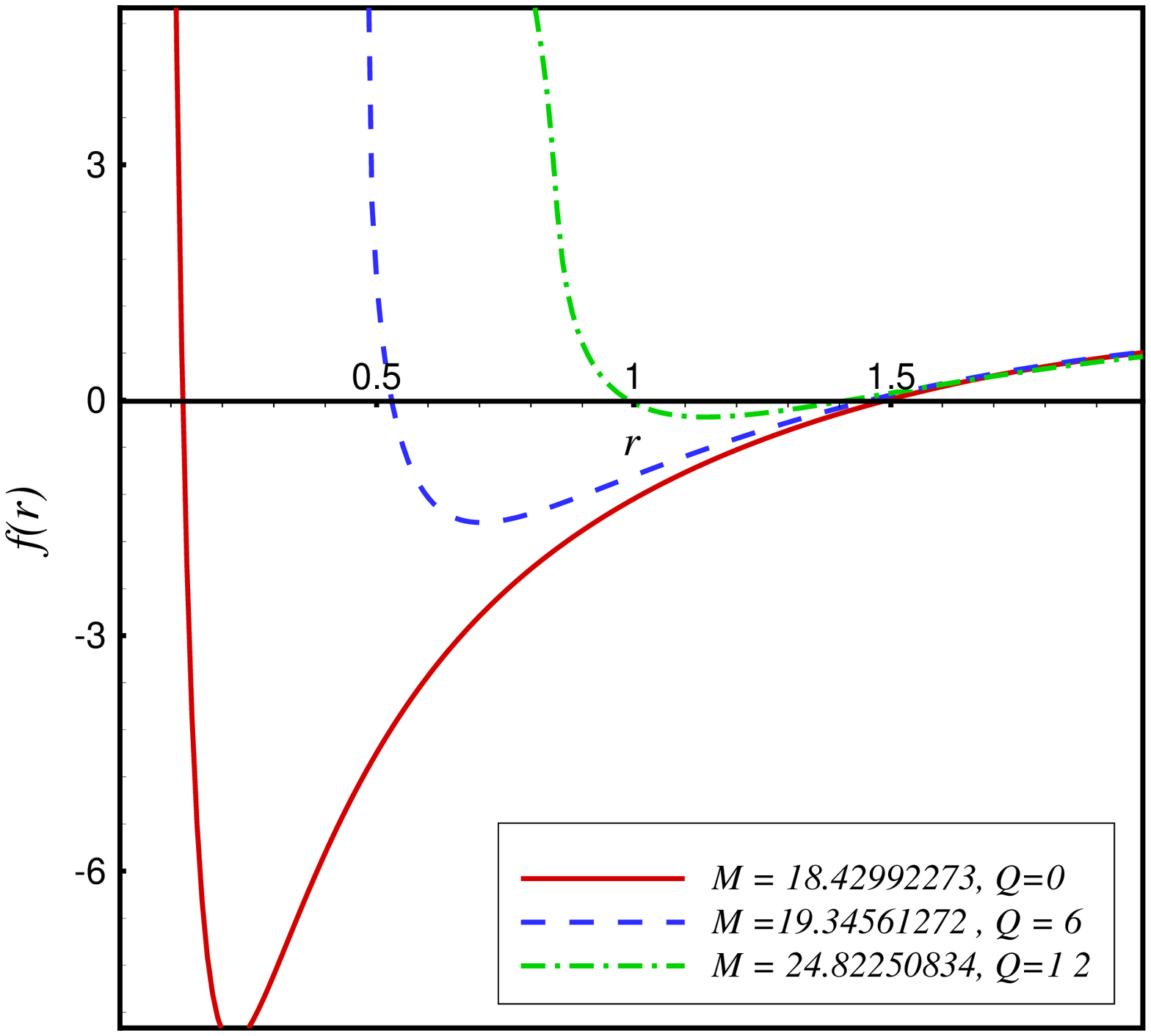}\label{fig5b}}\caption{Comparison of Einstein and FQT graviry for Asymptotically AdS solutions of Reissner-Nordstr\"om black hole with $k=1$, $\lambda=0.4$, $\mu=-0.1$ and $c=-0.0002$.}\label{fig5}
\end{figure}

\begin{figure}
\centering
\subfigure[Einsten Gravity]{\includegraphics[scale=0.27]{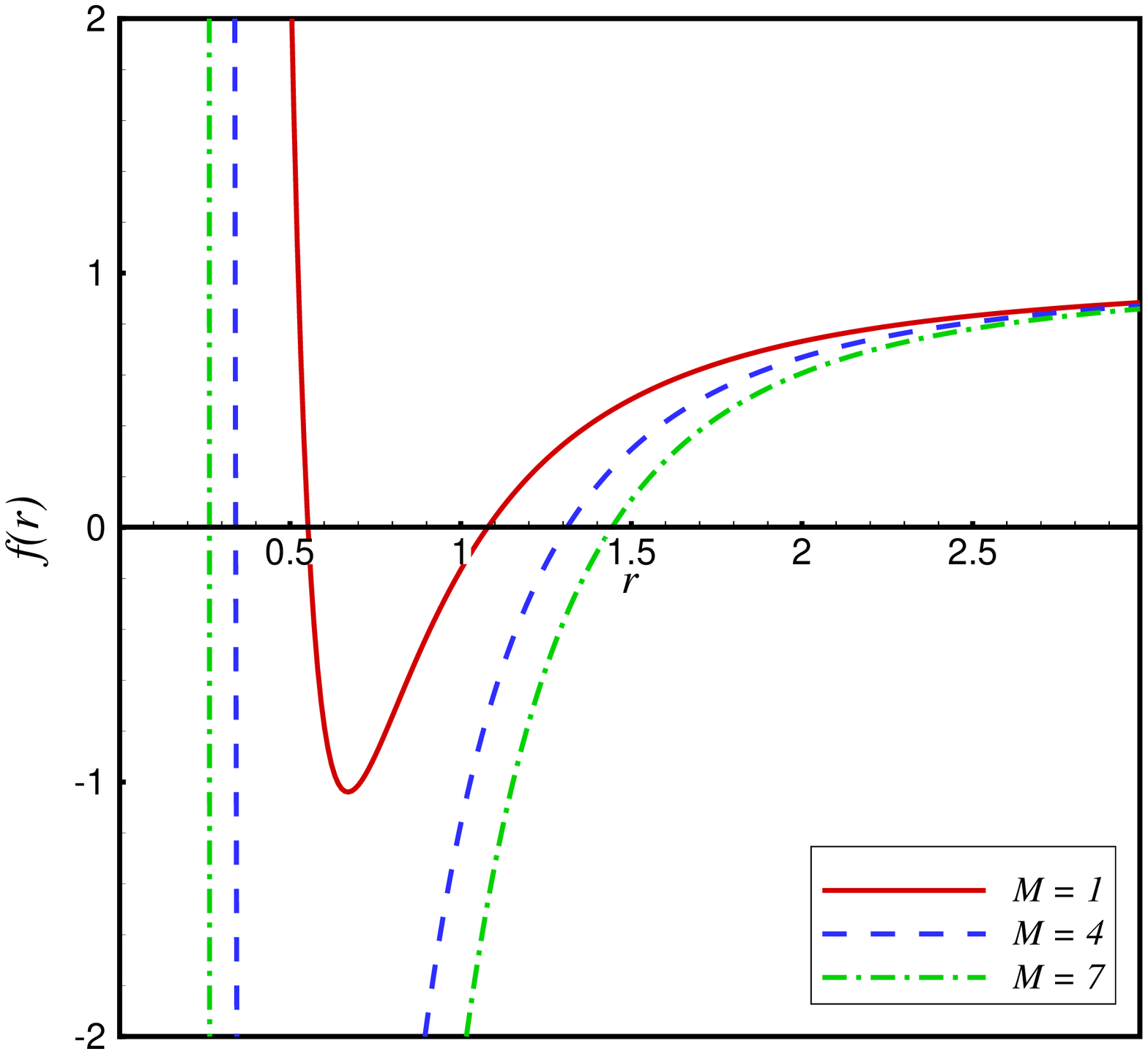}\label{fig6a}}\hspace*{.2cm}
\subfigure[TQT Gravity]{\includegraphics[scale=0.27]{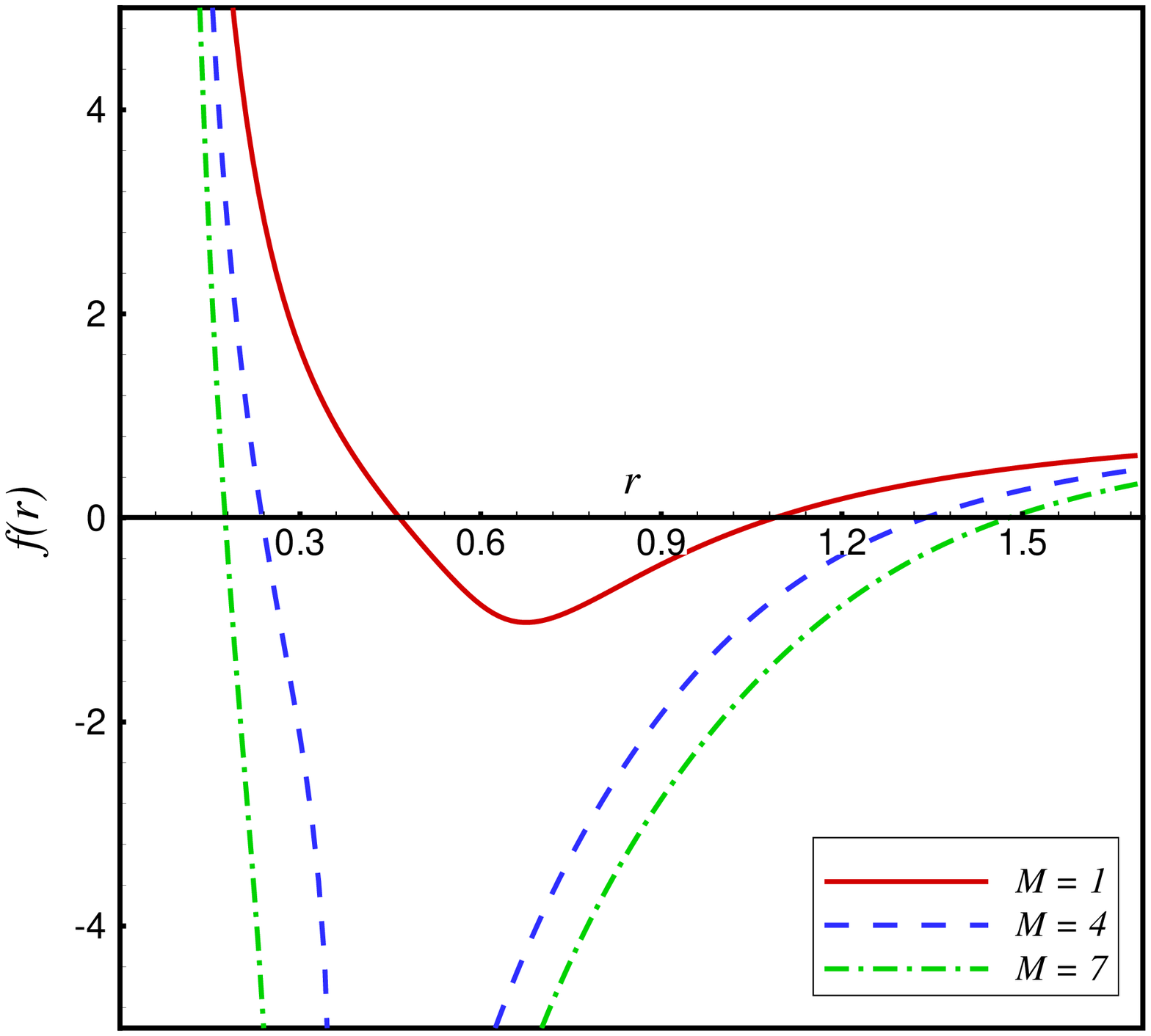}\label{fig6b}}\hspace*{.2cm}
\subfigure[FQT Gravity]{\includegraphics[scale=0.27]{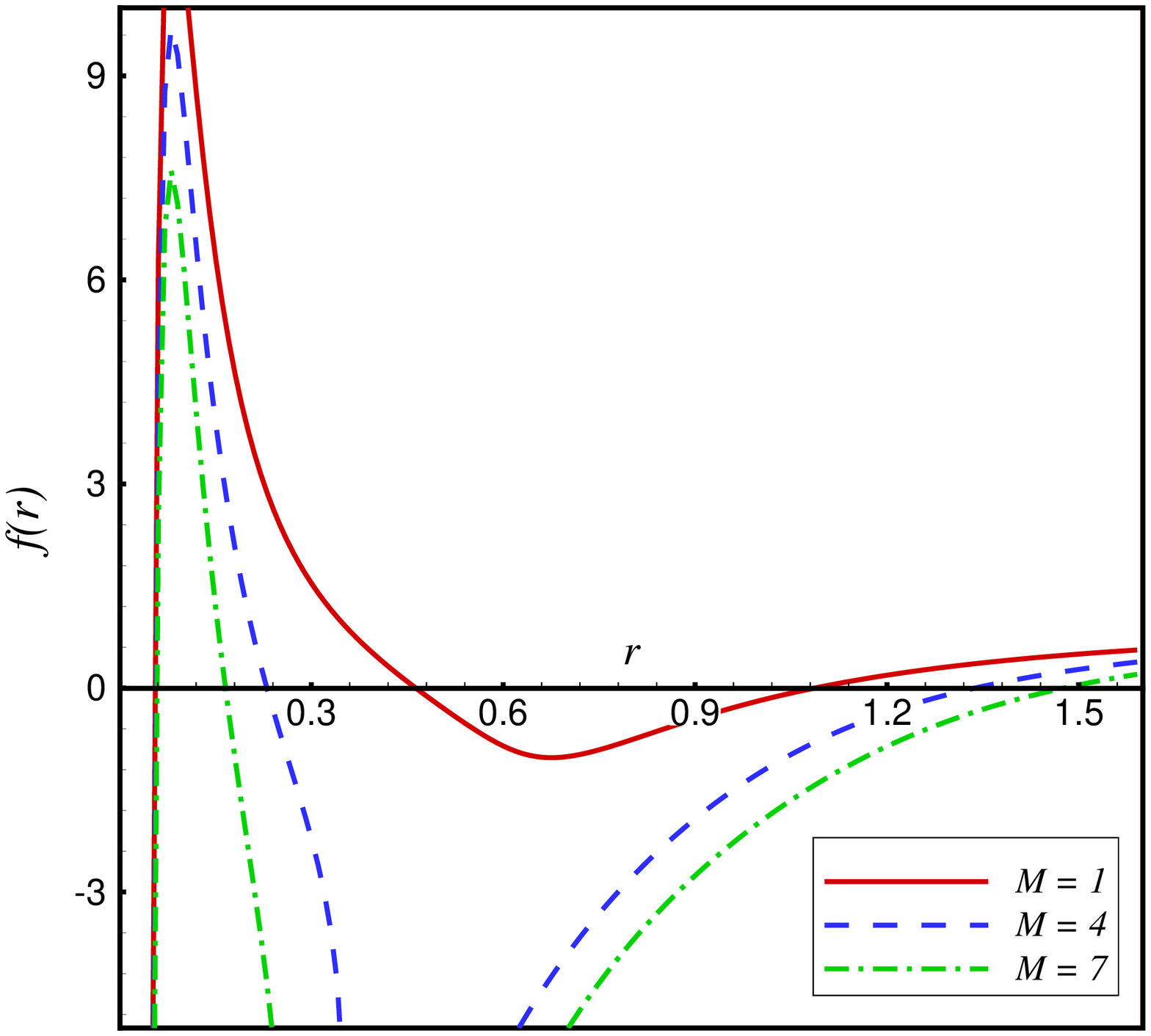}\label{fig6c}}\caption{Comparison of Einstein, TQT and FQT graviry for Asymptotically AdS solutions of Reissner-Nordstr\"om black hole with $Q=2$, $k=-1$, $\lambda=0.2$, $\mu=-0.1$ and $c=-0.0002$.}\label{fig6}
\end{figure}

\begin{figure}
\center
\includegraphics[scale=0.5]{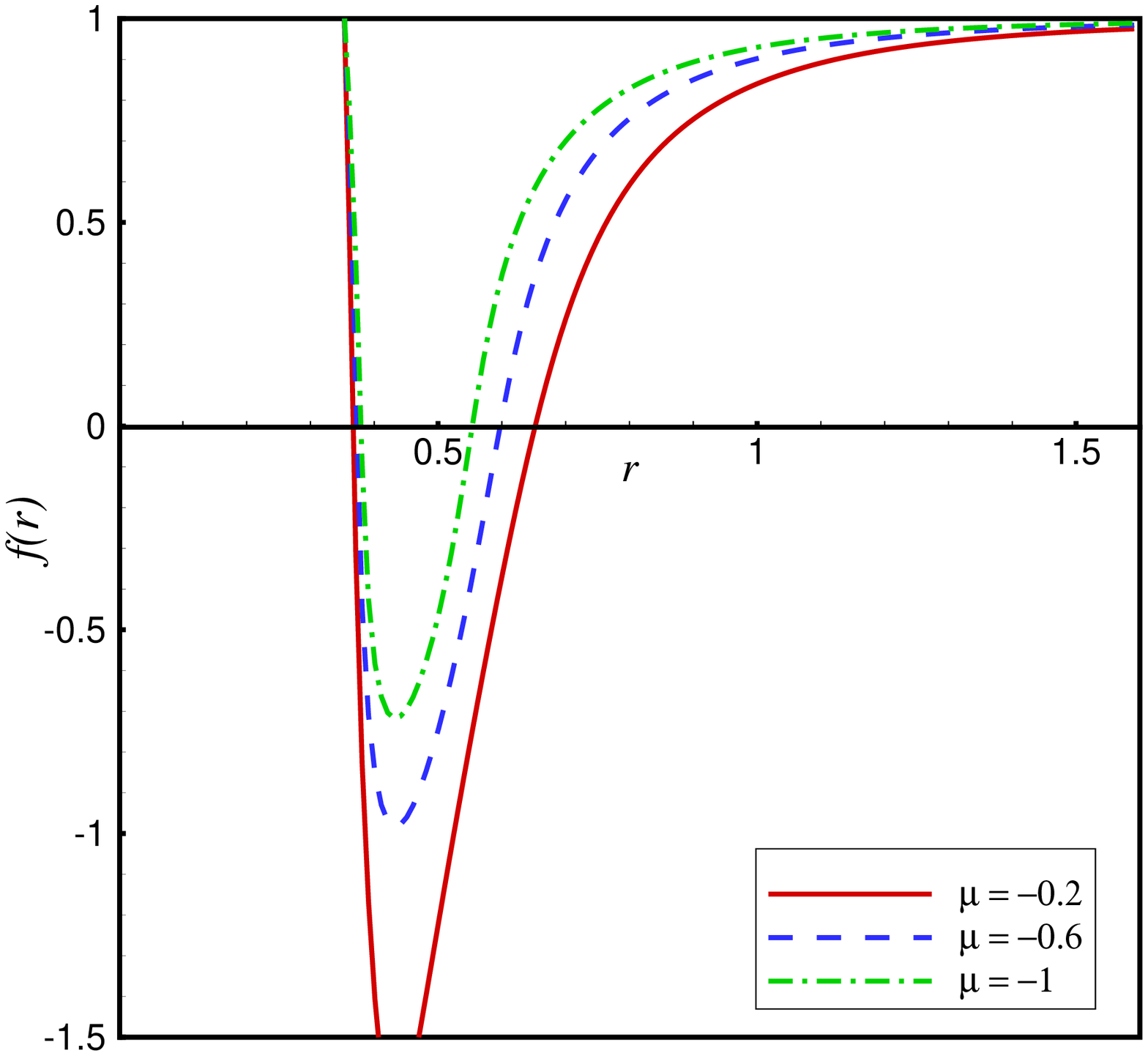}
\caption{\small{Asymptotically AdS solution $f(r)$ versus $r$ in FQT gravity with $M=1$, $Q=1$, $k=0$, $\lambda=-0.001$ and $c=-0.1$.} \label{fig7}}
\end{figure}

As one might expect, the boundary conditions at infinity ensure that the asymptotic symmetry group is the AdS group. Applying the limit $\lim_{r\rightarrow \infty} f(r)=1$ in Eq. (\ref{equasli}), we arrive at a modified definition for cosmological constant
\begin{equation}
\Lambda=-\frac{6(1-\lambda-\mu-c)}{L^2}.
\end{equation}
As the negative constant is one of the salient features of  AdS spacetimes, condition $\lambda+\mu+c<1$ ensures the cosmological constant to be negative.

 \ \
We can consider Reissner-Nordstr\"om solutions for a given black hole radius that we have plotted asymptotically AdS solution $f(r)$ versus r in Figs. (\ref{fig1})-(\ref{fig7}).
In Fig. (\ref{fig1}), we have investigated $f(r)$ for different value of $Q$ in FQT gravity. It shows that, there are two $Q_{\rm {min}}$ and $Q_{\rm {ext}}$. For fixed value of parameters $M$, $k$, $\lambda$, $\mu$ and $c$, depending to the value of $Q$, we can have a non extreme black hole for $Q <Q_{\rm {min}}$, a black hole with two horizons for $Q_{\rm {min}}<Q<Q_{\rm {ext}}$, an extremal black hole for $Q=Q_{\rm {ext}}$ and a naked singularity for $Q>Q_{\rm {ext}}$. In this figure, extremal black hole happens for $Q=10.23$.
Fig. (\ref{fig2}) shows that for fixed value of other parameters, there are  $M_{\rm {min}}$ and $M_{\rm {ext}}$ that depend on the value of $M$, we have a non extreme black hole, a black hole with two horizons, an extremal black hole or a naked singularity.

In Fig. (\ref{fig3}), we can see that by decreasing the value of $k$, the number of the roots of $f(r)$ (the horizons) becomes more. For example, for the fixed parameters $Q$, $M$, $\lambda$, $\mu$ and $c$, the function $f(r)$ has three horizons for $k=-1$. This can show the benefits and excellence of FQT gravity to other gravities because in this gravity we can have black holes with three horizons which is rare in other gravities. we should say that the third horizon is event horizon and the two other horizons are before it. As we don't have any knowledges about the inside of the black holes, so we can not speak about these two horizons.

Fig. (\ref{fig4}) shows that for $M=Q=0$, we have just one horizon ($r_{+}=1$) while as $Q$ and $M$ becomes larger, the number of horizons becomes more.

In Fig. (\ref{fig5}), we have compared the behavior of $f(r)$ in Einstein gravity and FQT gravity. As we know, One of the solutions of Einstein equation is Schwarzschild black hole. Schwarzschild black hole has only one horizon and its electric charge is 0 ($Q=0$). This is clear in Fig. (\ref{fig5a}) which there are one horizon for $Q=0$ and two horizons for $Q\neq 0$. But as Fig. (\ref{fig5b}) shows, in FQT gravity, there are two horizons for $Q=0$ which is in contrast with Einstein theory. This shows that, even if there is no charge in FQT gravity, FQT theory can have the effect of charge in $f(r)$. This is the best clear feature of FQT theory which distinguishes it from Einstein gravity.

In Fig. (\ref{fig6}), we have probed the behavior of $f(r)$ in Einstein gravity, TQT gravity and FQT gravity. In contrast with Einstein gravity and TQT gravity that have two horizons, there are three horizons in FQT gravity for $k=-1$. This can be the other priority of FQT gravity to both Einstein and TQT gravity.

In Fig.\eqref{fig7}, we have studied the influence of the coefficient of TQT gravity $(\mu)$ on f(r). It is clear that for other fixed parameters, there are two horizons where the first one is fixed and the second one increases as the parameter $\mu$ approaches to 0.



\subsection*{Asymptotically de Sitter spacetimes}
\begin{figure}
\centering
\subfigure[Einstein Gravity]{\includegraphics[scale=0.27]{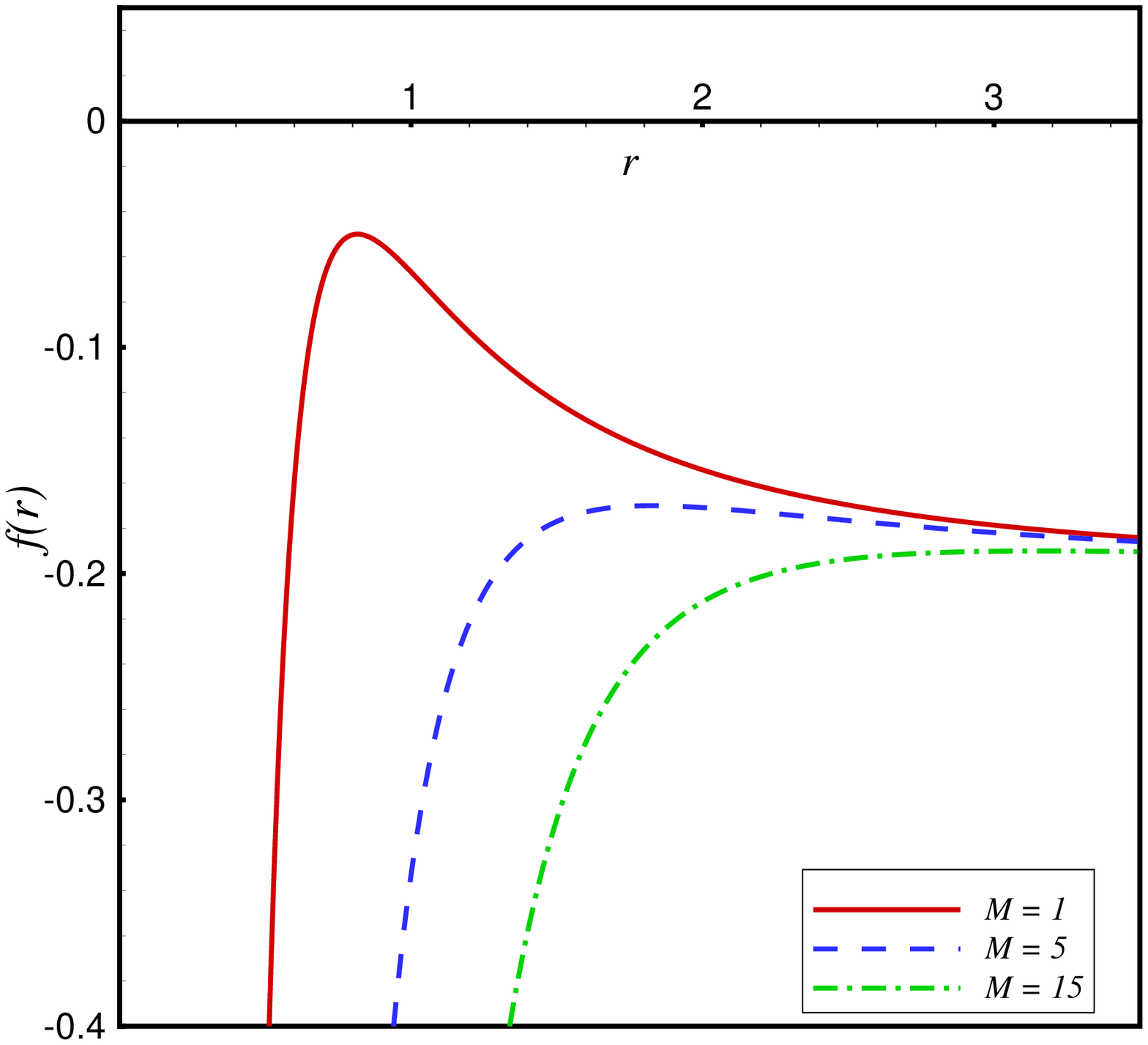}\label{fig8a}}\hspace*{.2cm}
\subfigure[FQT Gravity]{\includegraphics[scale=0.27]{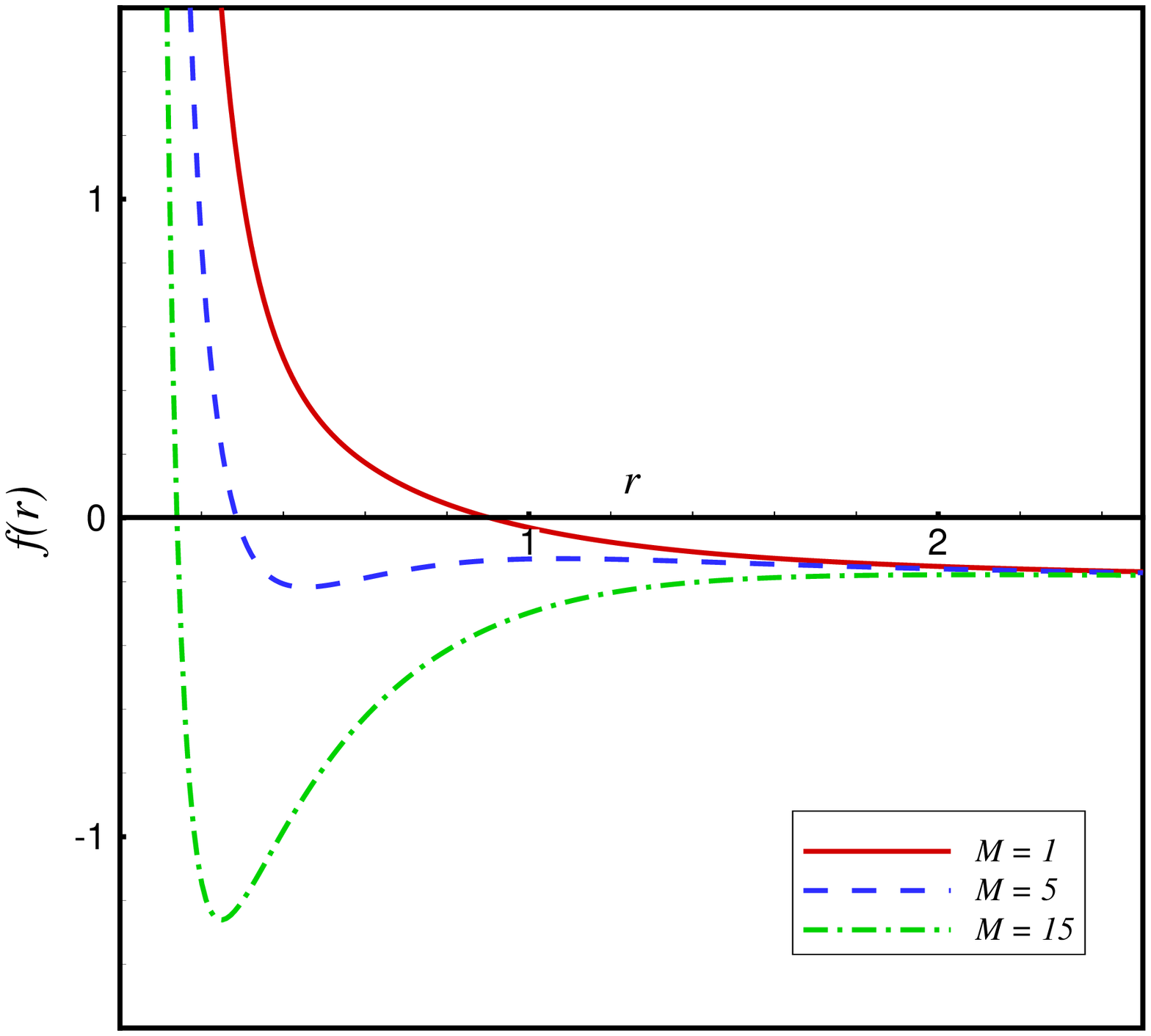}\label{fig8b}}\caption{Asymptotically dS solution $f(r)/5$ versus r with $Q=0$, $k=1$, $\lambda=0.4$, $\mu=-0.1$ and $c=-0.0002$.}\label{fig8}
\end{figure}

\begin{figure}
\centering
\includegraphics[scale=0.5]{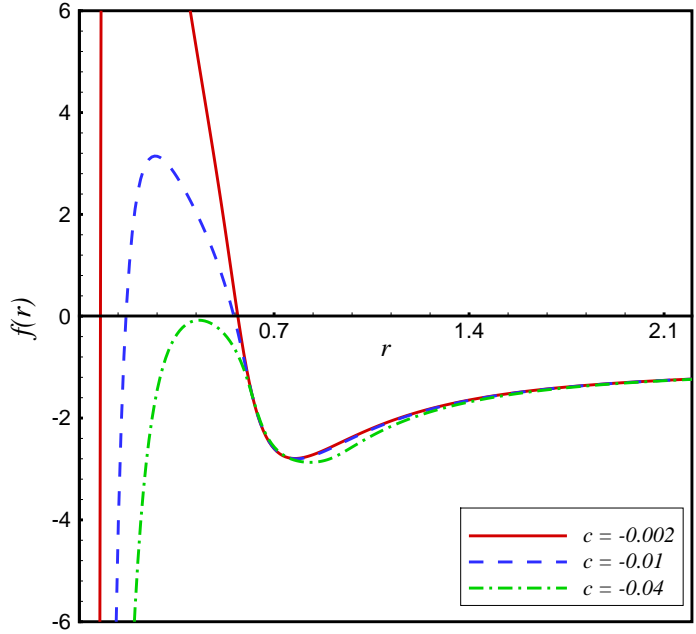}
\caption{\small{Asymptotically dS solution $f(r)$ versus $r$ in FQT gravity with $M=2$, $Q=3$, $k=-1$, $\lambda=-0.04$ and $\mu=-0.001$.} \label{fig9}}
\end{figure}

We have compared $f(r)$ in Einstein gravity and FQT gravity for asymptotically dS in figure (\ref{fig8}) and (\ref{fig9}) .
A spacetime is called asymptotically dS if it satisfies Einstein’s vacuum equation with a positive cosmological constant $\Lambda$. In order to guarantee this condition, we exert the constrain $\lim_{r\rightarrow \infty}f(r)=-1$. So the cosmological constant getting the form
\begin{equation}
\Lambda=\frac{6(1-\mu+\lambda+c)}{L^2}.
\end{equation}
It is clear from Fig. (\ref{fig8}) that although there is a naked singularity and no black hole in Einstein gravity, FQT gravity predicts a black hole with one horizon for the given parameters. So this gravity can show some black holes that is deniable in Einstein gravity.

The effects of the coefficient of FQT gravity on $f(r)$ have been shown in Fig. (\ref{fig9}). For $k=-1$ and fixed value of the other parameters, there are two black holes each with two horizons for $c>-0.04$ and an extreme black hole with $c=-0.04$. So negative parameter $c$ with small value can lead to a black hole with two horizons that their outer horizons are fixed but their inner ones become larger as parameter $c$ becomes smaller.



\subsection*{Asymptotically flat spacetimes}
\begin{figure}
\centering
\subfigure[Einstein Gravity]{\includegraphics[scale=0.27]{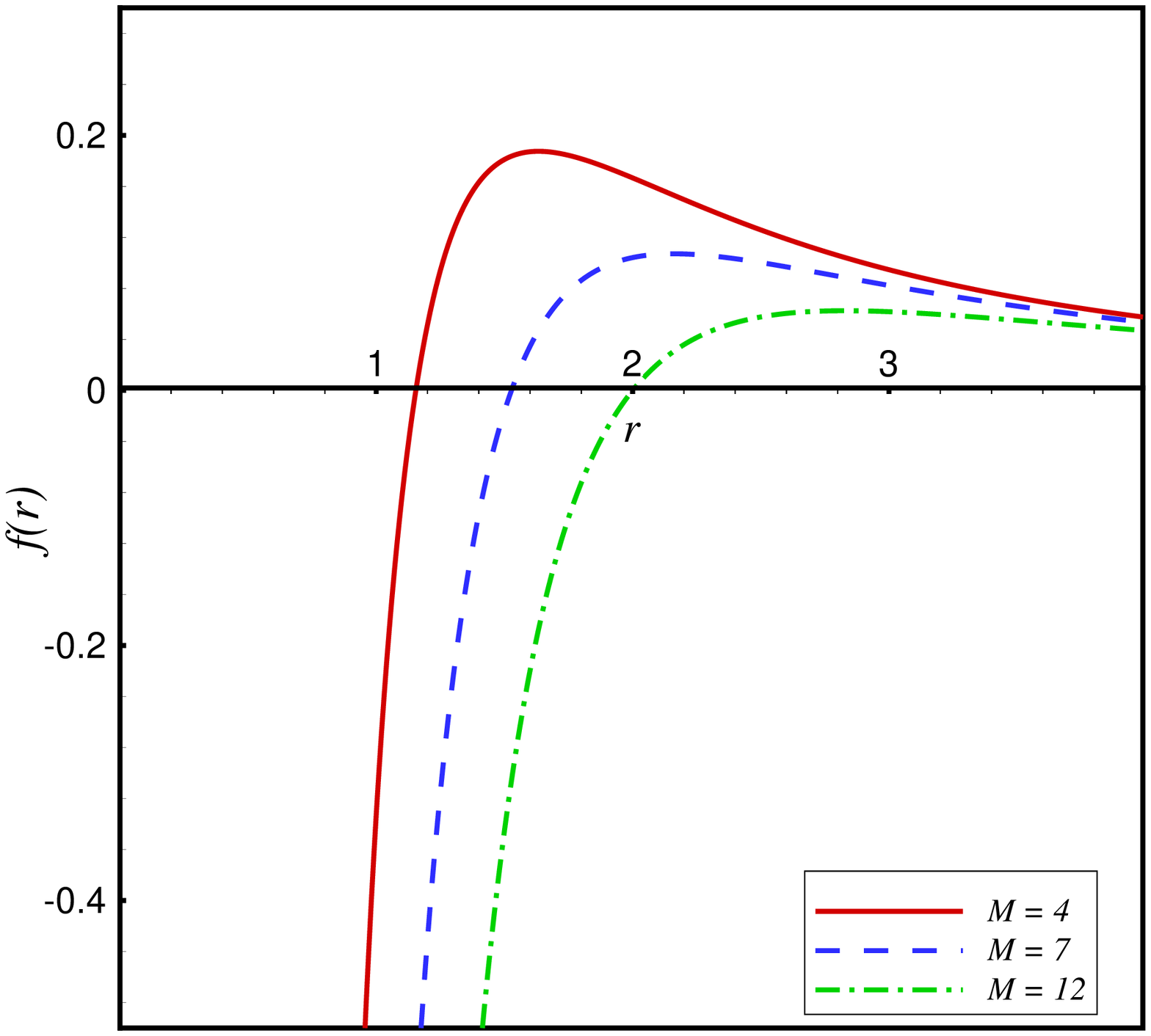}\label{fig10a}}\hspace*{.2cm}
\subfigure[FQT Gravity]{\includegraphics[scale=0.27]{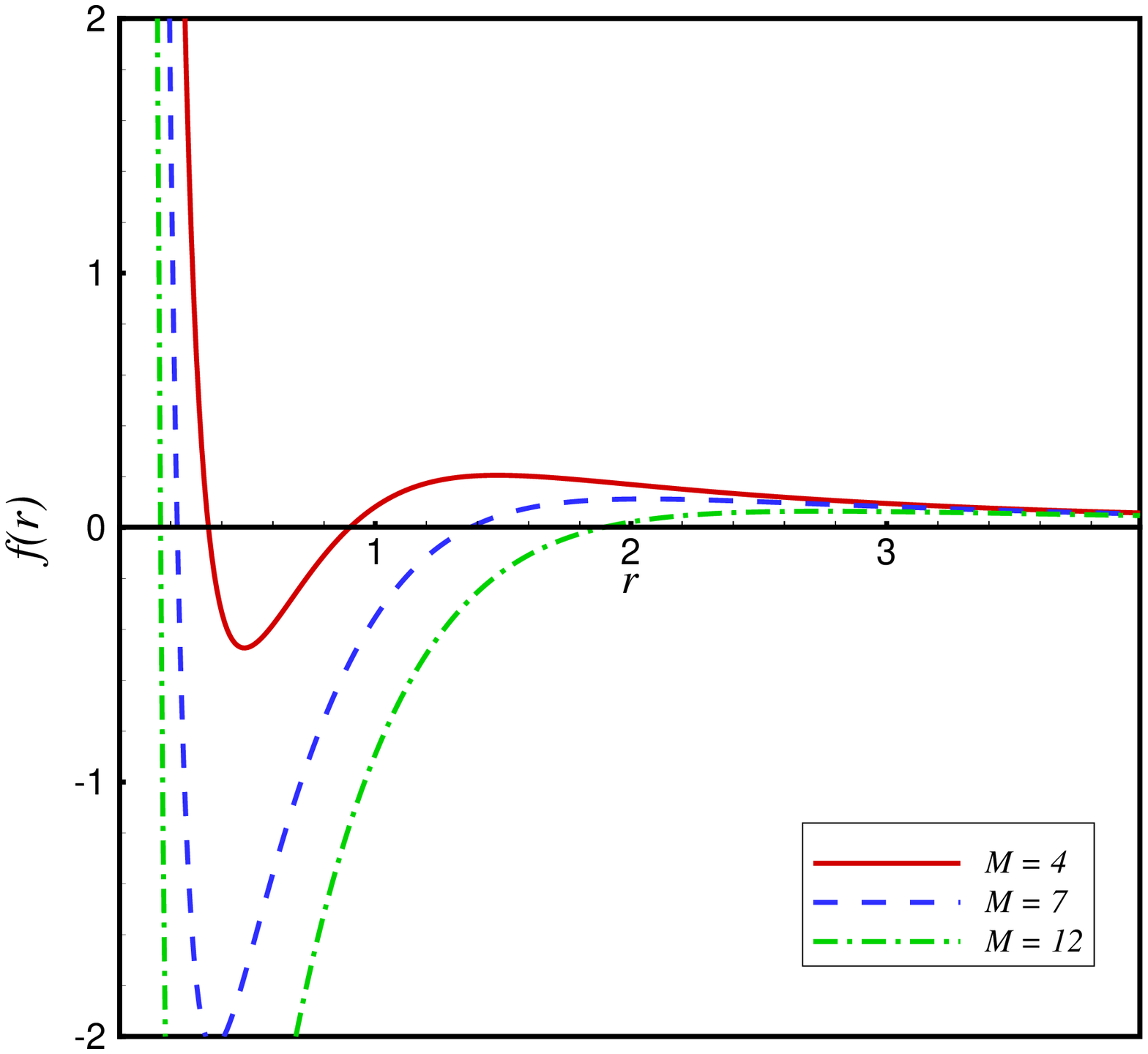}\label{fig10b}}\caption{Asymptotically flat solution $f(r)$ versus $r$ with $Q=0$, $k=1$, $\lambda=0.4$, $\mu=-0.1$ and $c=-0.0002$.}\label{fig10}
\end{figure}

Fig. (\ref{fig10}) shows the comparison of $f(r)$ between Einstein and FQT gravity in asymptotically flat space.
Heuristically, asymptotically flat spacetimes are spacetimes that approach Minkowski space at “large distances” from some spacetime region. To achieve this goal, we use $\lim_{r\rightarrow \infty}f(r)=0$ which leads to set $\Lambda=0$. The metric function $f(r)$  behaves asymptotically flat at large distances $f(r)\sim \frac{L^2}{r^2}$. Comparing the roots of $f(r)$ in Einstein and FQT theory one by one shows that for these parameters, the values of roots in FQT are smaller than the values of roots in Einstein gravity. This shows that for these parameters, the black hole in FQT is smaller than the one in Einstein gravity. We know that whatever the radius of the horizons is smaller, the black hole is usually more stable. So the black holes in FQT gravity are usually more stable than the ones in Einstein gravity.

\subsection*{Thermal stability}
In this section, we would like to study thermal stability of the solutions. We can study the stability of a thermodynamic system like a black hole by investigating the behavior of energy $m(S,q)$ with respect to small variations of thermodynamic coordinates $S$ and $q$. To have the local stability, $m(S,q)$ should be a convex function of its extensive variables. For this purpose, we use Hessian matrix in grand ensemble in which two extensive parameters $S$ and $q$ are changing. Hessian matrix is
\begin{eqnarray}
H=\left[
\begin{array}{ccc}
H_{11} & H_{12}\\
H_{21} & H_{22}
\end{array} \right],
\end{eqnarray}
where
\begin{equation}
H_{11}=\Big(\frac{\partial ^2 m}{\partial S^2}\Big)=\frac{2r^2}{9\pi L^2A^3}\{BA-CAr+BDr\}\,\,,\,\,
H_{22}=\Big(\frac{\partial ^2 m}{\partial q^2}\Big)=\frac{8\pi}{r^2}\,\,,\,\,
H_{12}=H_{21}=\Big(\frac{\partial ^2 m}{\partial S\partial q}\Big)=-\frac{64\pi q r}{3A}\,\,,\,\,
\end{equation}
and
\begin{eqnarray}
A=4ck^3L^6-3\mu k^2L^4r^2+2\lambda kL^2r^4+r^6,
\end{eqnarray}
\begin{eqnarray}
B=64 q^2 \pi^2 L^2 r^2-6 \mu_{0} r^8+6 c k^4 L^8-3\mu k^3 L^6 r^2-3 k L^2 r^6,
\end{eqnarray}
\begin{eqnarray}
C=128 q^2 \pi^2 L^2 r-48 \mu_{0} r^7-6\mu k^3 L^6 r-18 k L^2 r^5,
\end{eqnarray}
\begin{eqnarray}
D=-6\mu k^2 L^4 r+8 k \lambda L^2 r^3 +6 r^5.
\end{eqnarray}

\begin{figure}
\centering
\subfigure{\includegraphics[scale=0.27]{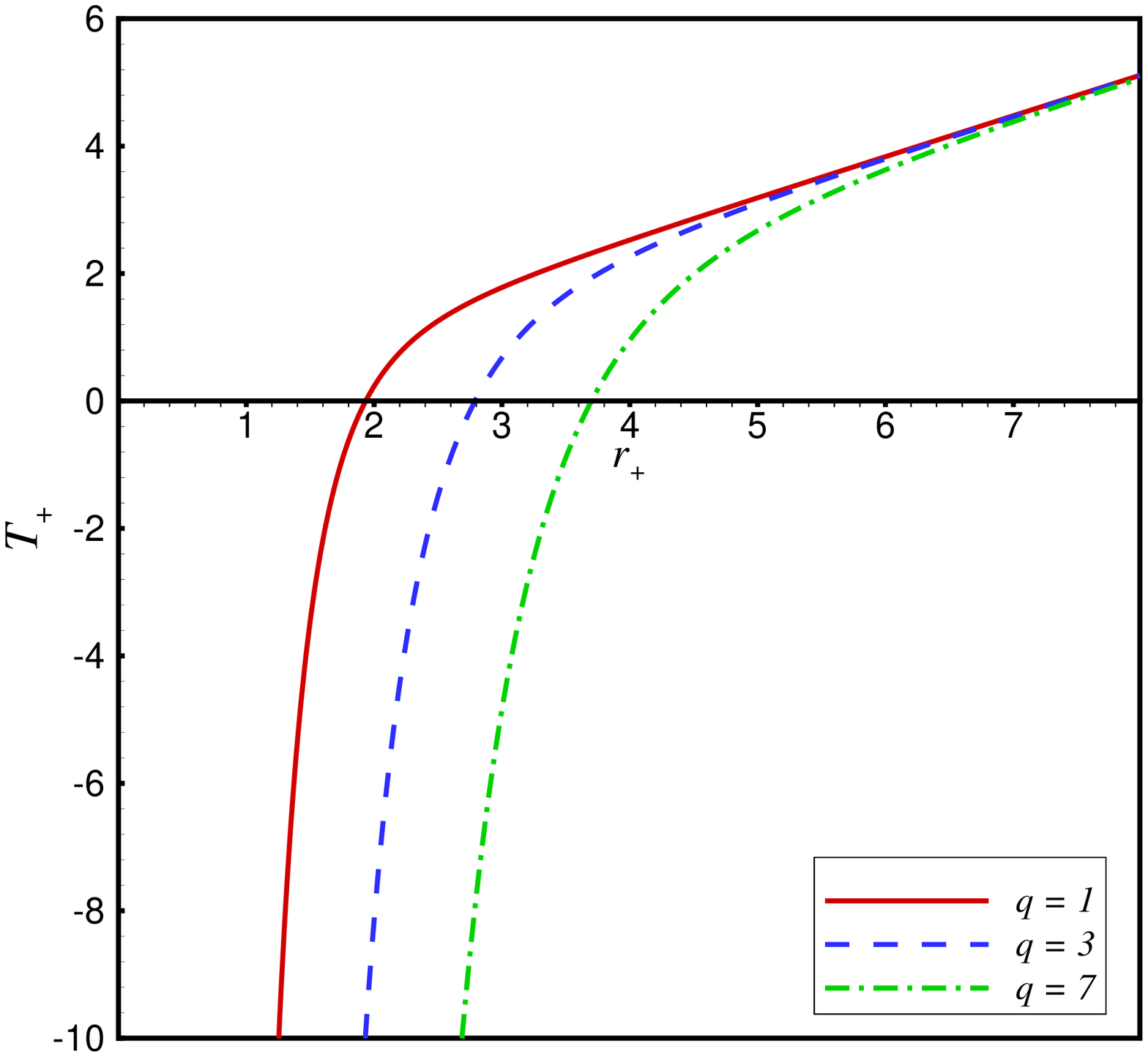}\label{fig11a}}\hspace*{.2cm}
\subfigure{\includegraphics[scale=0.27]{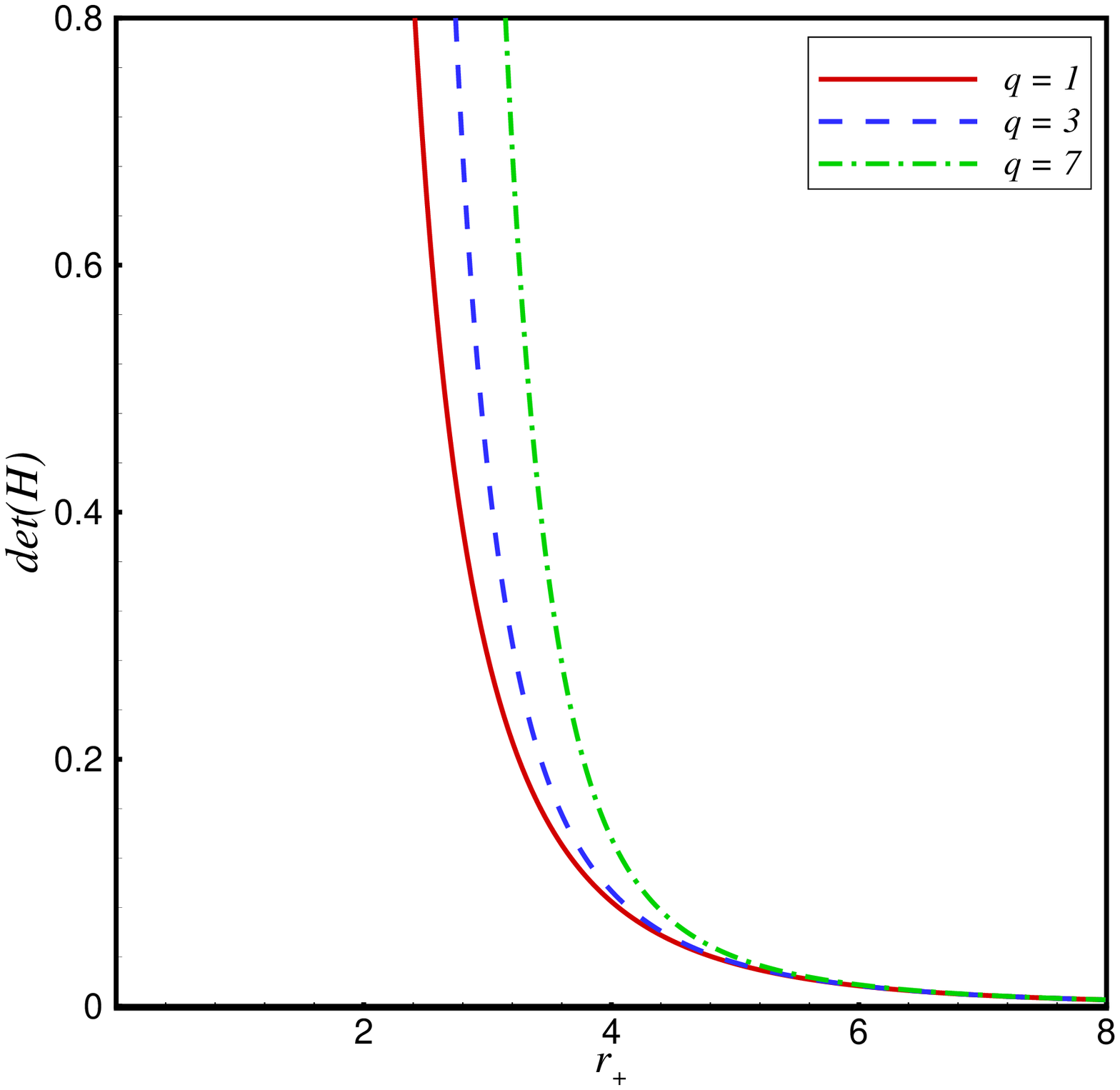}\label{fig11b}}\caption{Temperature and det(H) versus $r_{+}$ for AdS solution with $k=0$, $\lambda=-0.01$, $\mu=-0.5$ and $c=-0.5$.}\label{figure11}
\end{figure}

\begin{figure}
\centering
\subfigure{\includegraphics[scale=0.27]{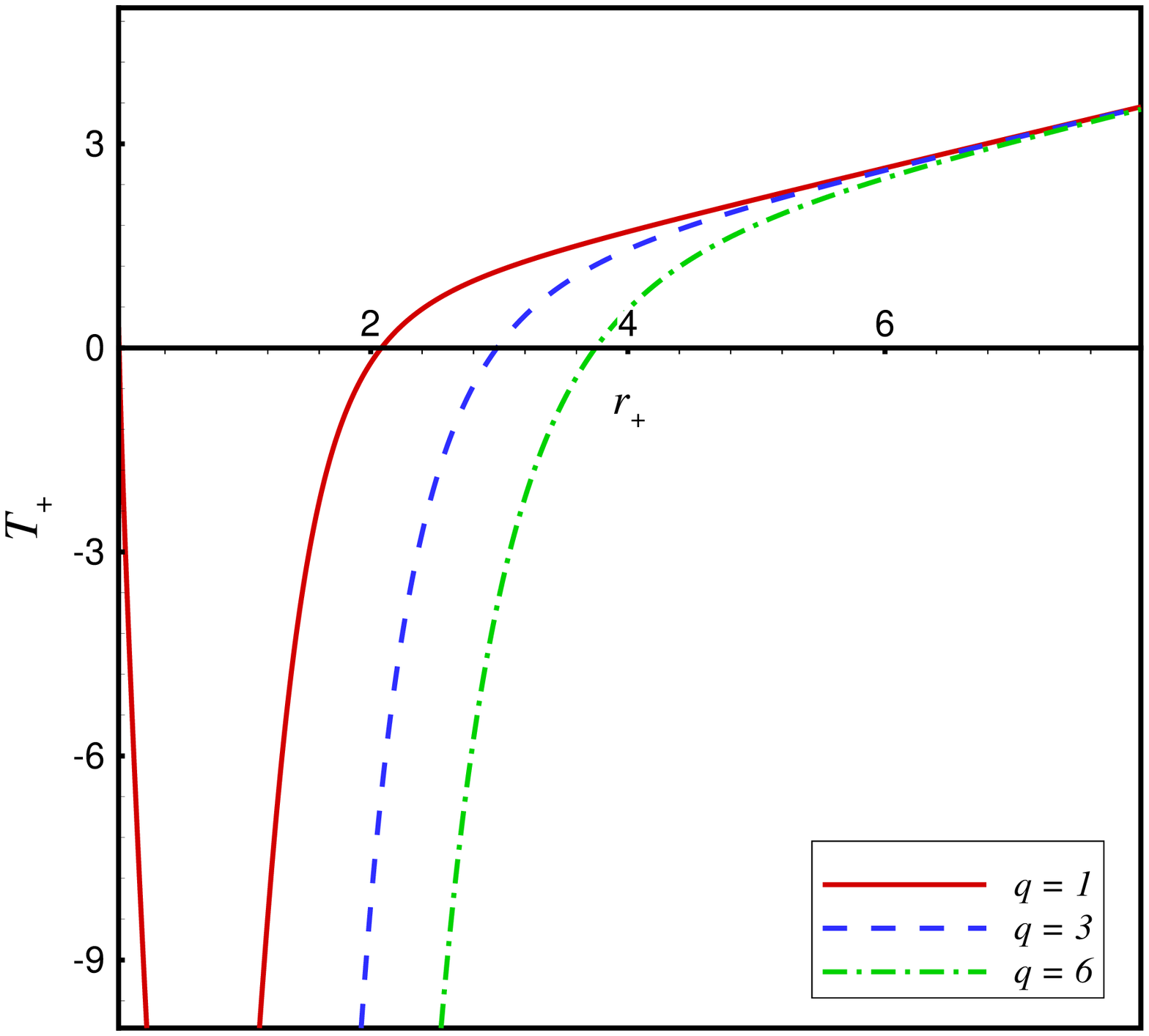}\label{fig12a}}\hspace*{.2cm}
\subfigure{\includegraphics[scale=0.27]{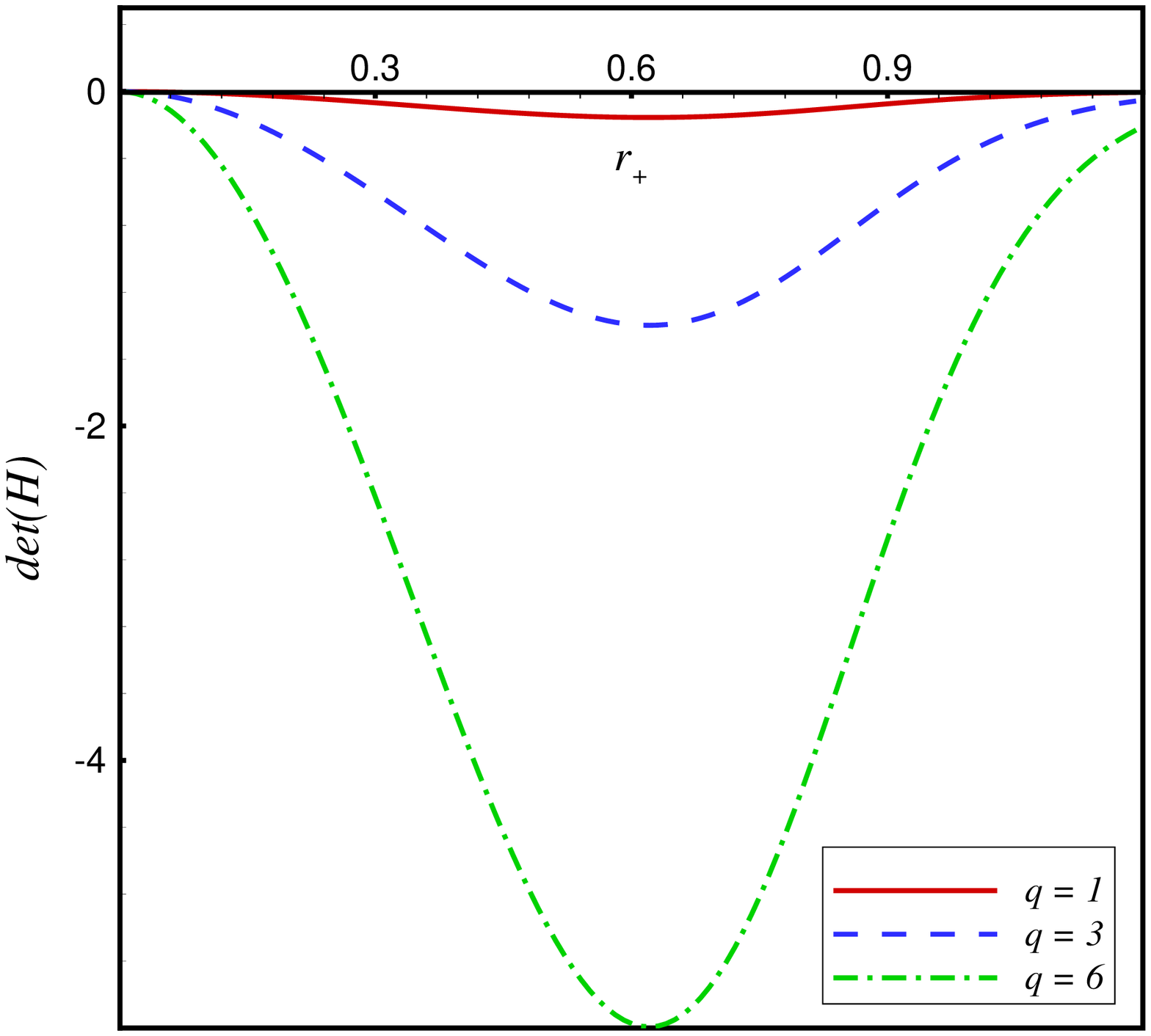}\label{fig12b}}\hspace*{.2cm}
\subfigure{\includegraphics[scale=0.27]{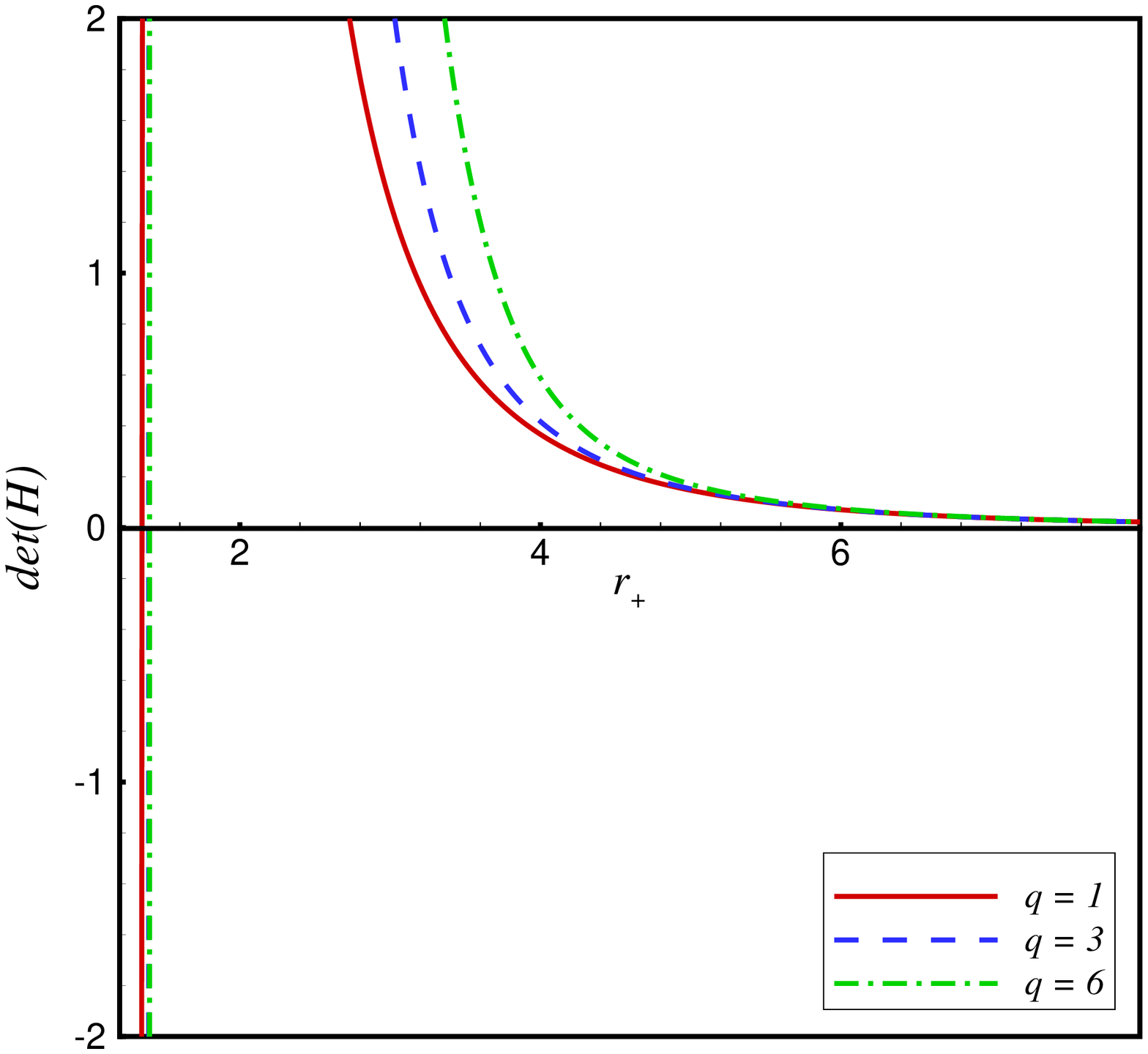}\label{fig12c}}\caption{Temperature and det(H) versus $r_{+}$ for AdS solution with $k=-1$, $\lambda=-0.002$, $\mu=-0.2$ and $c=-0.2$.}\label{figure12}
\end{figure}

\begin{figure}
\centering
\subfigure{\includegraphics[scale=0.27]{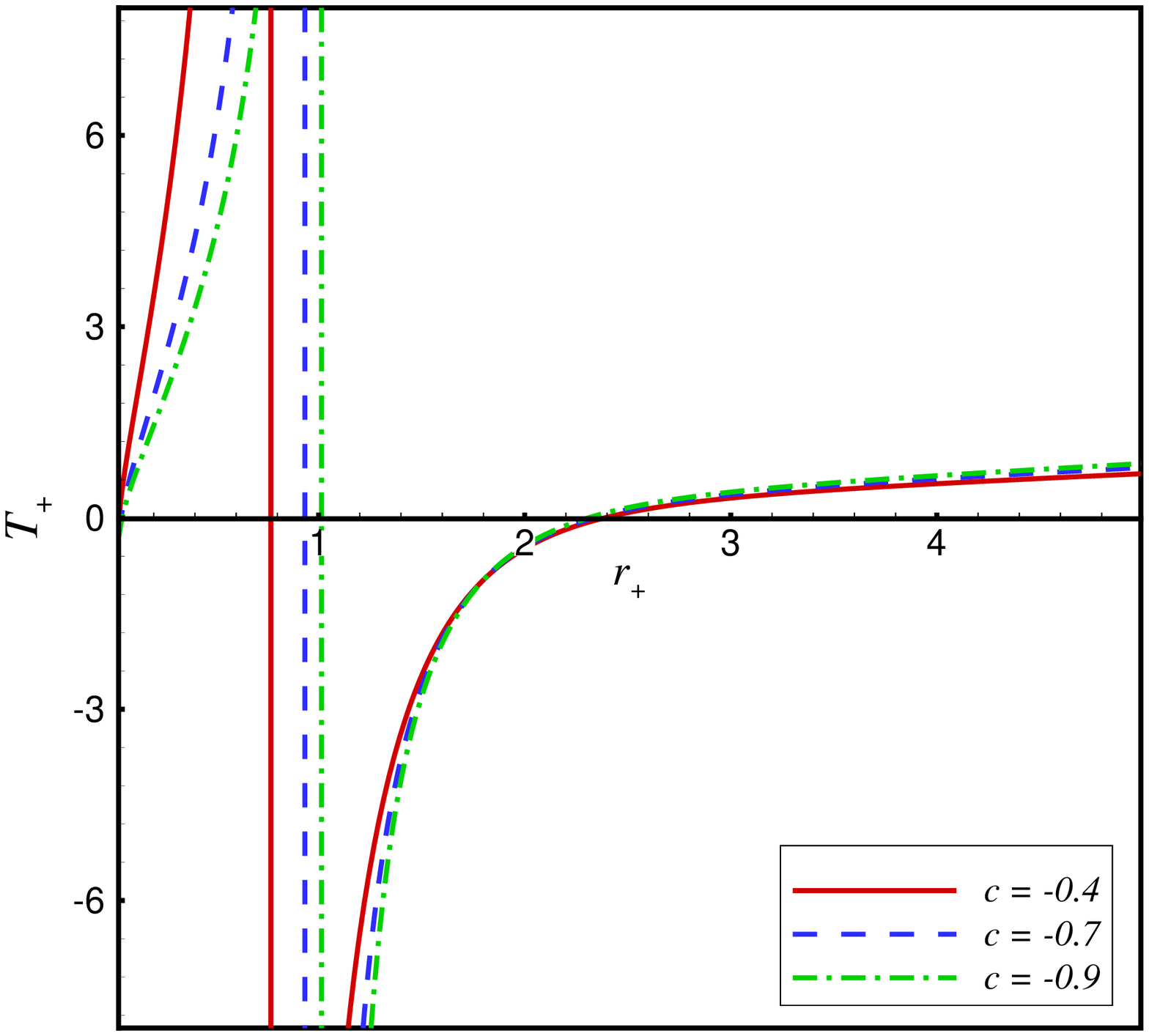}\label{fig13a}}\hspace*{.2cm}
\subfigure{\includegraphics[scale=0.27]{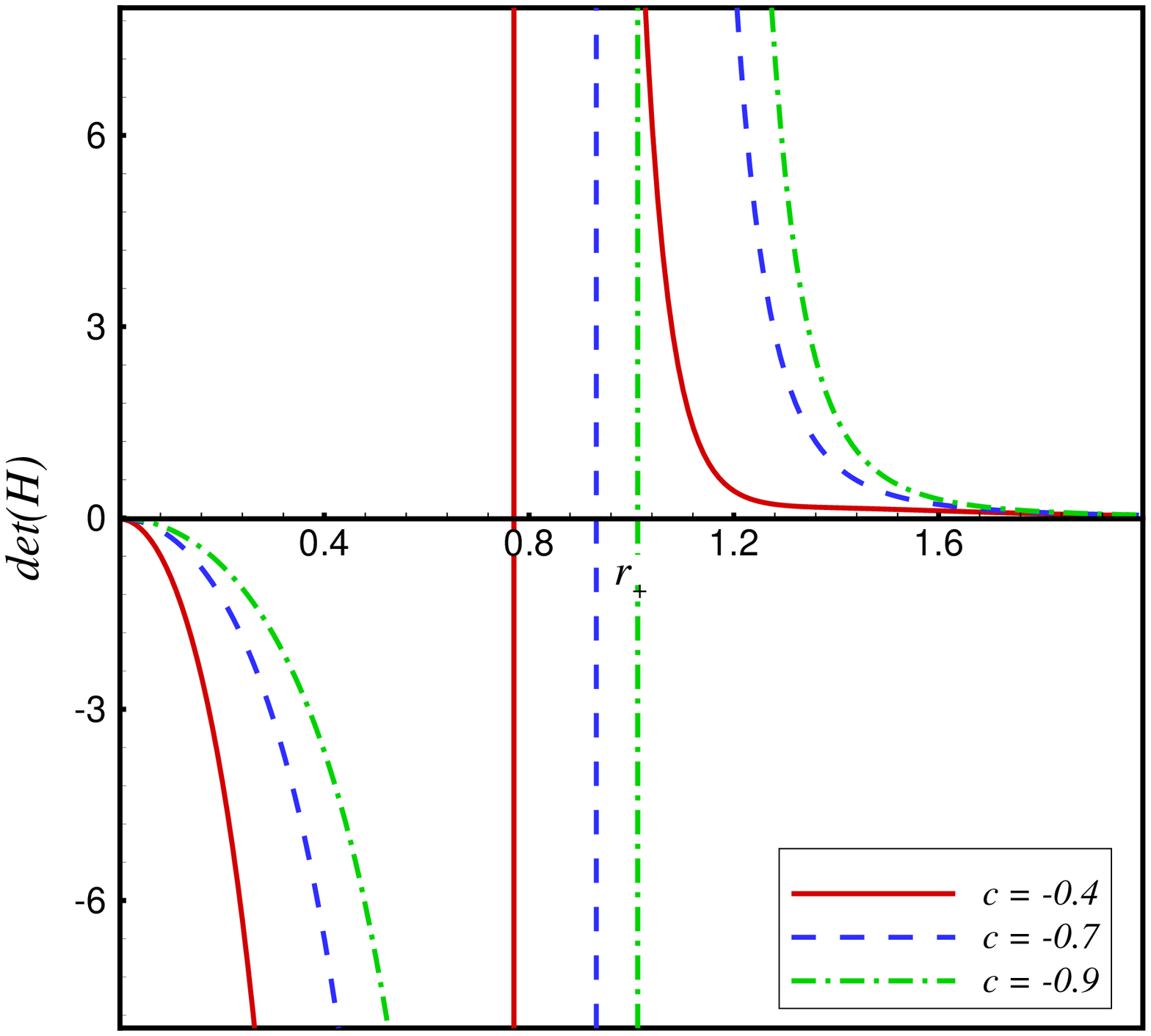}\label{fig13b}}\hspace*{.2cm}
\subfigure{\includegraphics[scale=0.27]{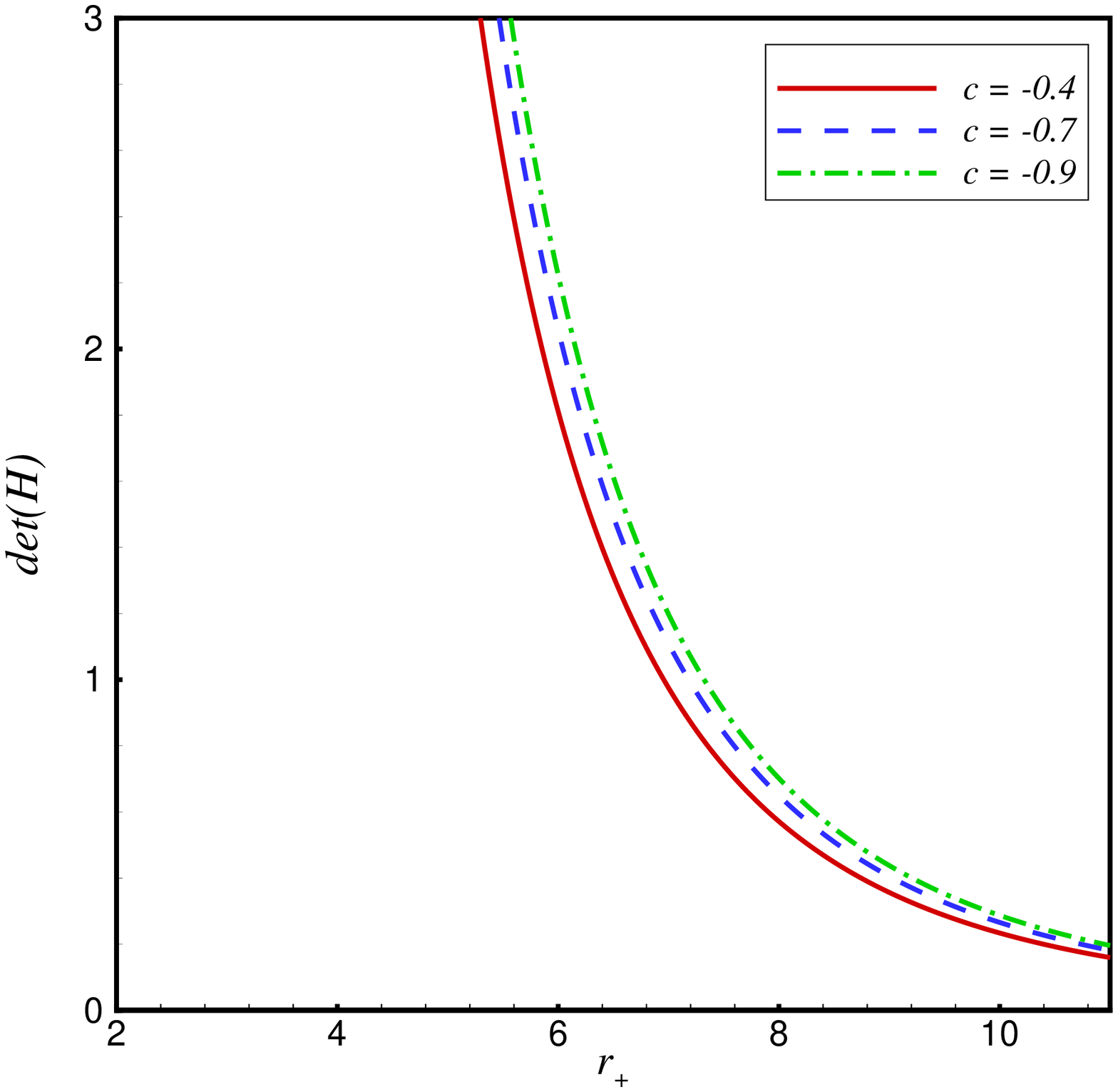}\label{fig13c}}\caption{Temperature and det(H) versus $r_{+}$ for AdS solution with $k=1$, $q=2$, $\lambda=0.002$ and $\mu=-0.8$.}\label{figure13}
\end{figure}

\begin{figure}
\centering
\subfigure{\includegraphics[scale=0.27]{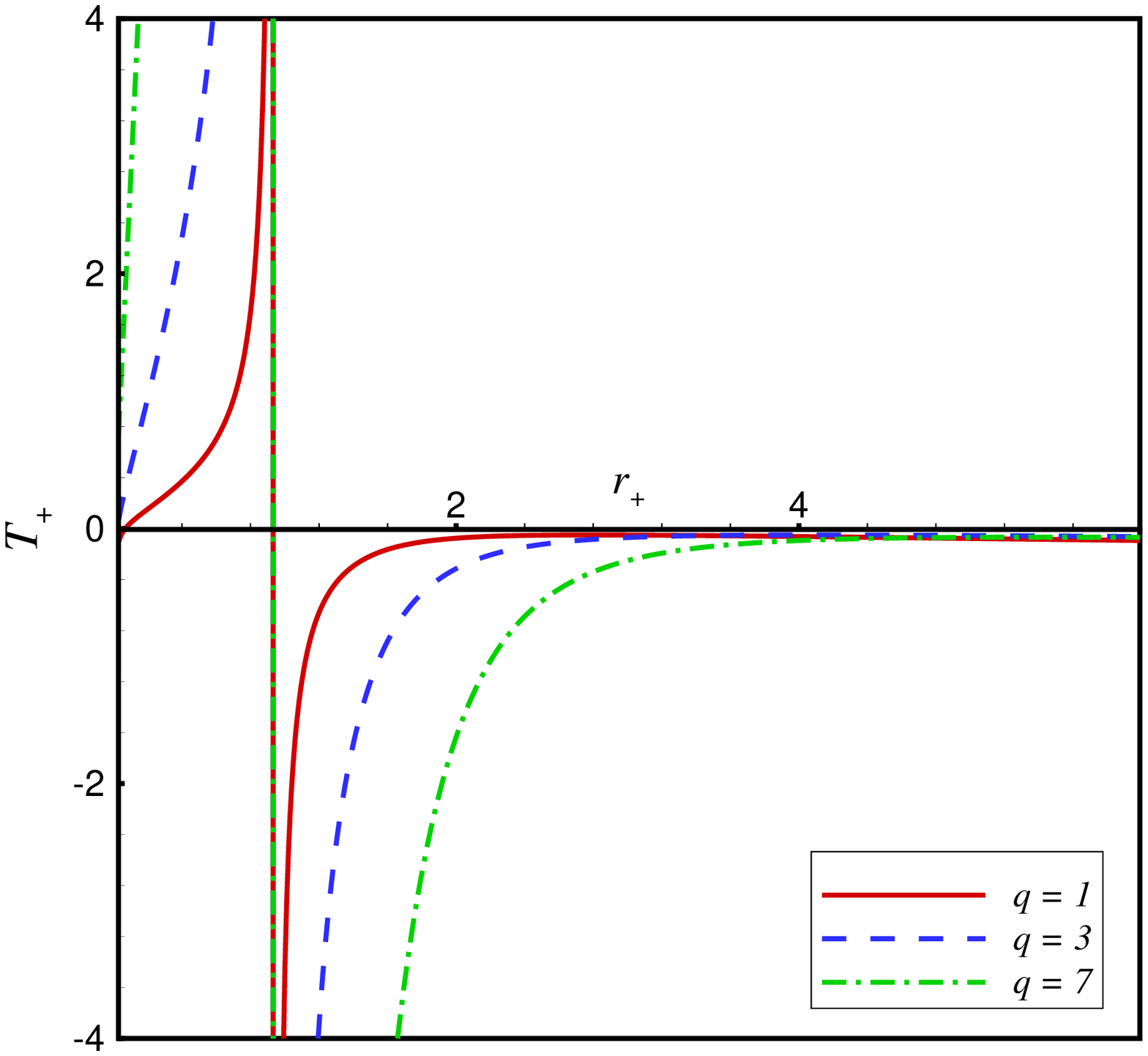}\label{fig14a}}\hspace*{.2cm}
\subfigure{\includegraphics[scale=0.27]{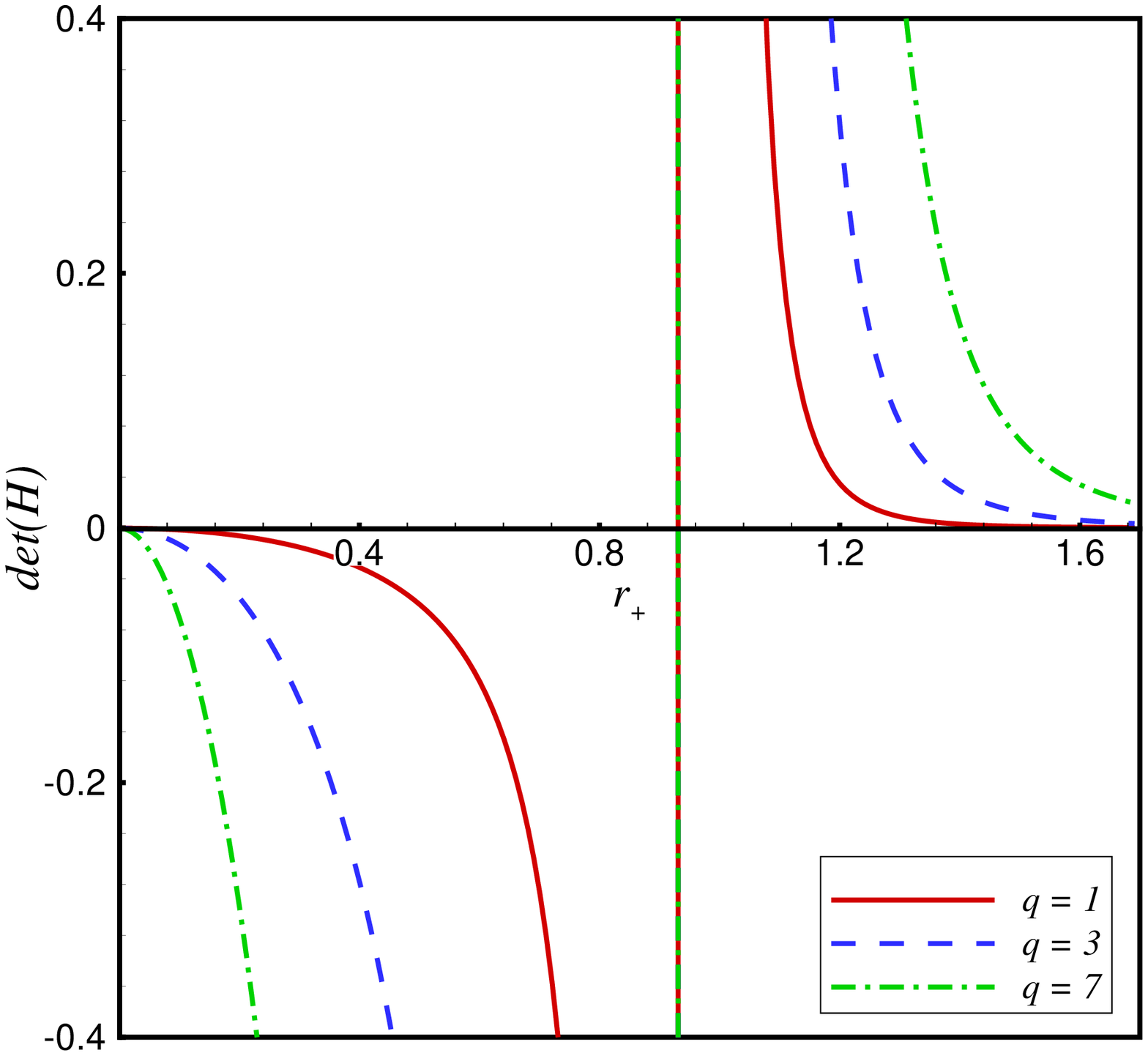}\label{fig14b}}\hspace*{.2cm}
\subfigure{\includegraphics[scale=0.27]{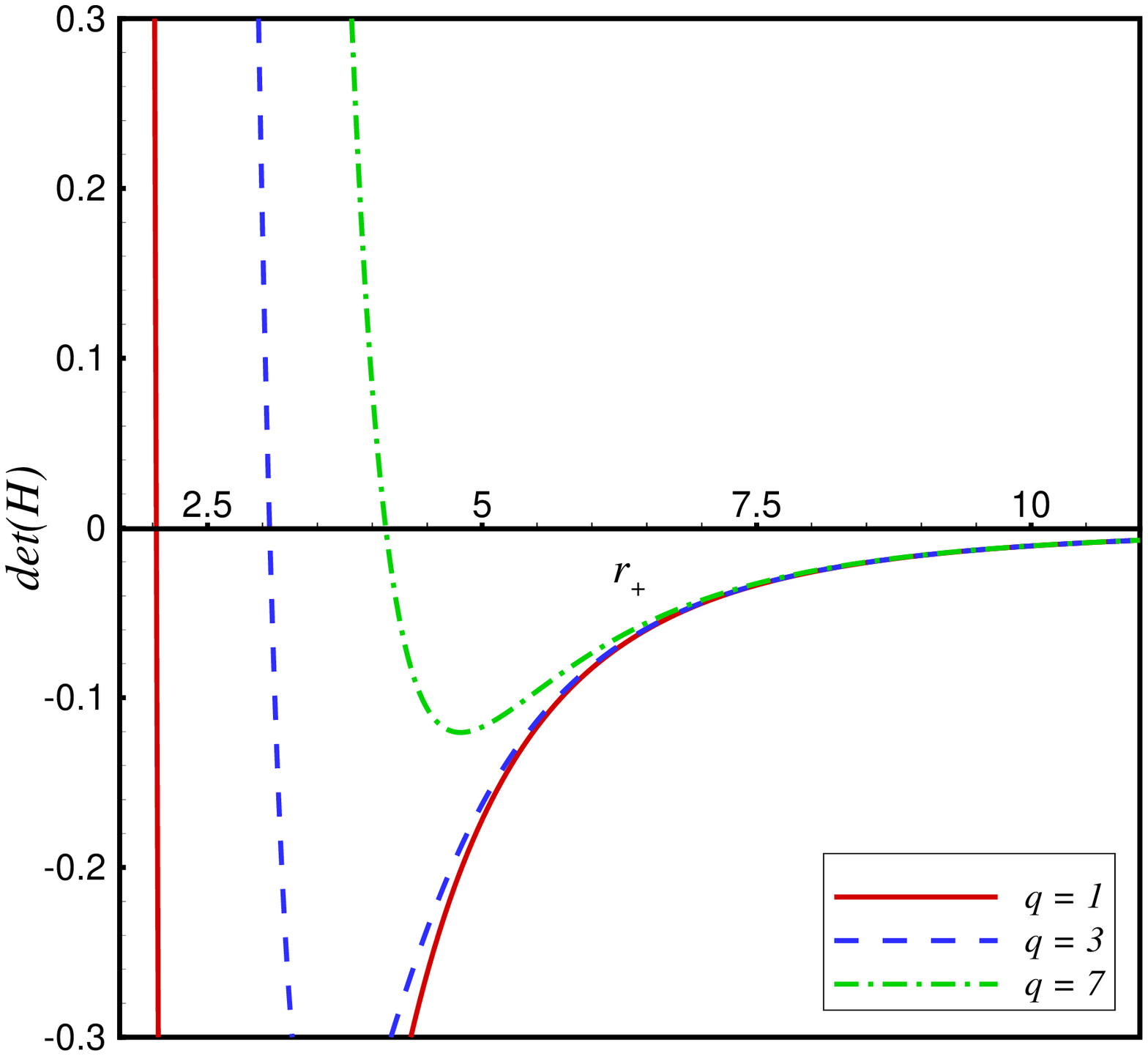}\label{fig14c}}\caption{Temperature and det(H) versus $r_{+}$ for dS solution with $k=1$, $\lambda=-0.01$, $\mu=-0.5$ and $c=-0.5$.}\label{figure14}
\end{figure}

\begin{figure}
\centering
\subfigure{\includegraphics[scale=0.27]{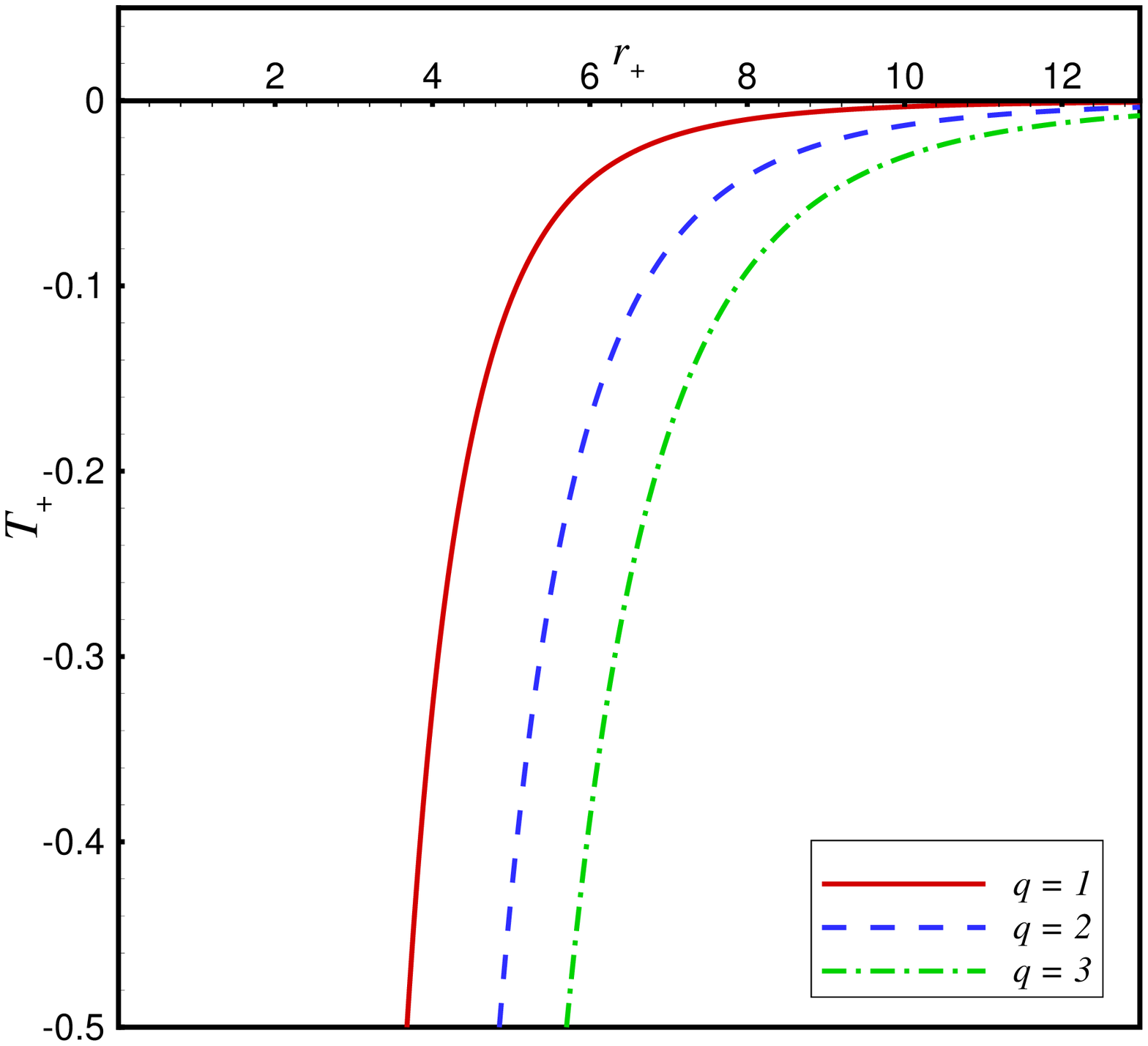}\label{fig15a}}\hspace*{.2cm}
\subfigure{\includegraphics[scale=0.27]{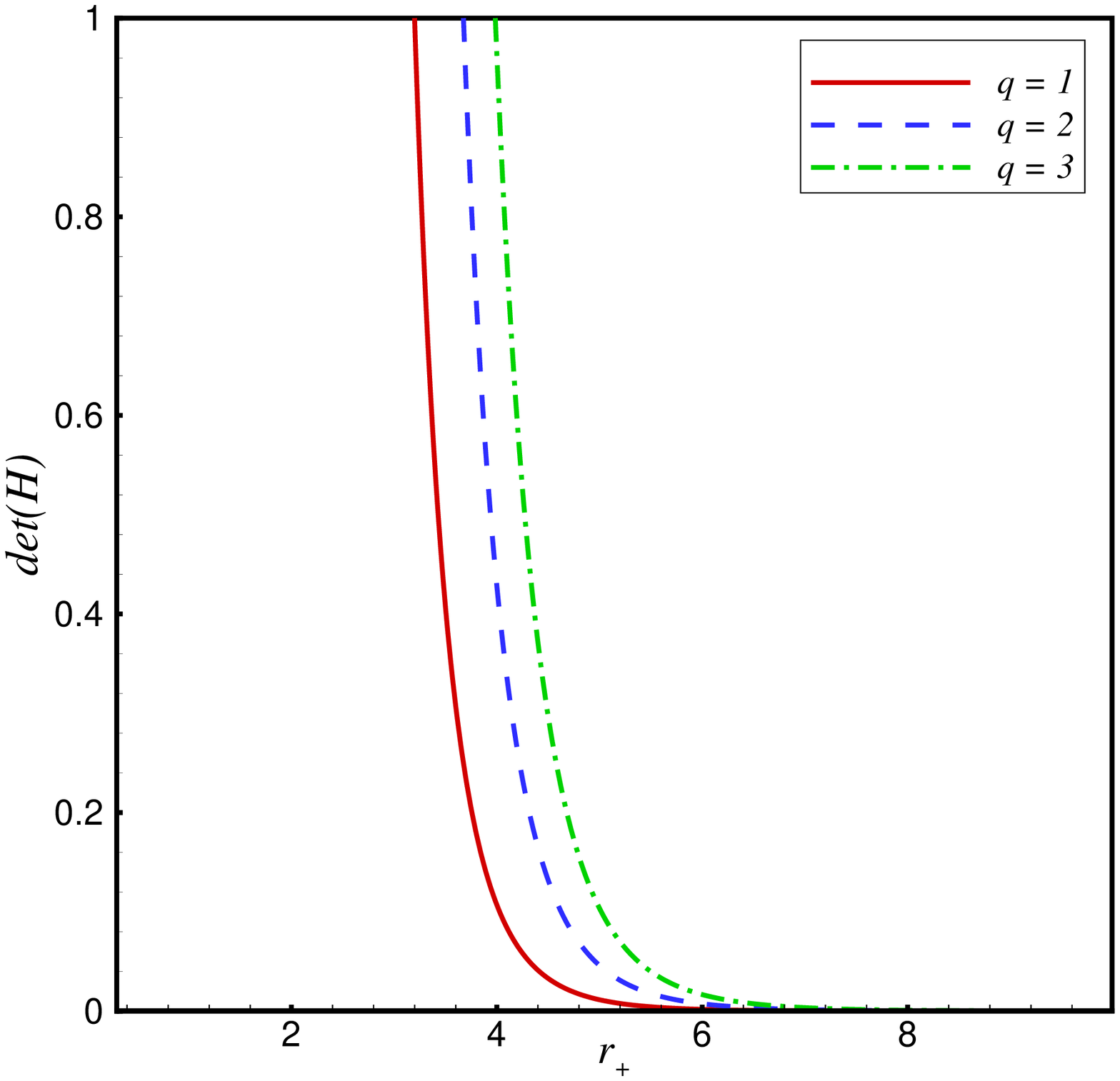}\label{fig15b}}\caption{Temperature and det(H) versus $r_{+}$ for asymptotically flat solution with $k=0$, $\lambda=0.002$, $\mu=-0.5$ and $c=-0.7$.}\label{figure15}
\end{figure}

Positive value for the determinant of Hessian matrix (we abbreviate it to det(H) for simplicity) Guarantees the stability of this black hole. On the other hand, negative temperature is not physical and so we should ignore them. Therefore, to have thermal stability for Reissner-Nordstr\"om black holes in FQT gravity, we should find the regions in which both det(H) and $T_{+}$ are positive simultaneously.\\
We have plotted figures \eqref{figure11}-\eqref{figure15} for $L=1$ to show the stability of this black hole in AdS, dS and flat solutions respectively. By changing the parameter $q$ in Fig. \eqref{figure11} for AdS solutions with $k=0$ and fixed parameters $\lambda$, $\mu$ and $c$, det(H) is positive for all value of $r_{+}$ even though it goes to 0. By this condition, the positive value of $T_{+}$ is the determinative for the stability. For temperature, there is a ${r_{+}}_{min}$ for each value of $q$ which $T_{+}$ is positive for $r_{+}>{r_{+}}_{min}$. By increasing the value of $q$, the value of ${r_{+}}_{min}$ increases, so small value for $q$ leads to a larger region for stability.

In Fig. \eqref{figure12}, we have repeated the stability of the black hole for different values of $q$ but for $k=-1$ and other fixed parameters. Unlike the positive value of det(H) for all $r_{+}$ in $k=0$, det(H) is only positive for $r_{+}>{r_{+}}_{1}$ in $k=-1$. ${r_{+}}_{1}$ is approximately the same for all $q$ and its value is less than 2. Disregard to the value of $k$, $T_{+}$ has a similar behavior like the one in the previous figure in which ${r_{+}}_{min}$ is more than 2 for all value of $q$. So for these parameters, the regions with positive temperature ($r_{+}>{r_{+}}_{min}$) lead to the stability for this black hole.

To see that how the coefficient of FQT gravity can effect the stability, we have plotted Fig. \eqref{figure13} for different values of $c$. we can see that for each value of $c$, det(H) has a singularity in ${r_{+}}_{s}$ where det(H) is positive for $r_{+}>{r_{+}}_{s}$ and negative for $r_{+}<{r_{+}}_{s}$. Increasing the parameter $c$ leads to a smaller value for ${r_{+}}_{s}$. There is also a singularity for temperature in ${r_{+}}_{b}$ which separates the behavior of $T_{+}$ for $r_{+}<{r_{+}}_{b}$ and $r_{+}>{r_{+}}_{b}$. $T_{+}$ is positive for $r_{+}<{r_{+}}_{b}$, but for $r_{+}>{r_{+}}_{b}$, it depends to the value of $r_{+}$ and is positive for $r_{+}>{r_{+}}_{c}\sim 2.4$. Comparing the behaviors of det(H) and $T_{+}$ shows that det(H) and $T_{+}$ are Simultaneously positive for $r_{+}\gtrsim 2.4$. So the stability for these parameters doesn't depend on the value of it is established if the value of $r_{+}$ is almost larger than 2.4.

Fig. \eqref{figure14} shows the behaviors of det(H) and $T_{+}$ for dS solutions and different value of $q$ with other fixed parameters. It shows that $T_{+}$ has a singularity in $r_{+}={r_{+}}_{d}$ which $T_{+}>0$ for $r_{+}<{r_{+}}_{d}$ and $T_{+}<0$ for $r_{+}>{r_{+}}_{d}$. There is also a singularity for det(H) in ${r_{+}}_{e}$ where $T_{+}$ is positive only for ${r_{+}}_{e}<r_{+}<{r_{+}}_{max}$. In ${r_{+}}_{max}$, det(H) changes to negative values from the positive ones. It is also clear that by increasing the value of $q$, ${r_{+}}_{max}$ increases. The obtained results show that positive det(H) and positive $T_{+}$ don't have any joint regions. So dS solutions don't have thermal stability.
we have shown Fig. \eqref{figure15} to certify that flat solutions like dS solutions don't demonstrate thermal stability.


\section{concluding results}
In this paper, we constructed the solutions of Reissner-Nordstr\"om black hole in the presence of quartic quasi topological gravity. This gravity contains terms quartic in the curvature and leads to second-order equations
of motion for an arbitrary space-time in each dimensions $n \geq 4$ except 8. The solution of this theory can lead to a new gravitational solution which is valid in physical theory by AdS/CFT correspondence.
We also obtained conserved and thermodynamic quantities. These quantities obeyed the first law of thermodynamic. Then, we investigated the physical properties of the solutions in three parts: AdS, dS and flat spacetime. Depending on the value of $Q$ and for fixed value of parameters like $M$, $k$, $\lambda$, $\mu$ and $c$, the solutions result to a non extreme black hole for $Q <Q_{\rm {min}}$, a black hole with two horizons for $Q_{\rm {min}}<Q<Q_{\rm {ext}}$, an extremal black hole for $Q=Q_{\rm {ext}}$ and a naked singularity for $Q>Q_{\rm {ext}}$.\\
In the absence of electric charge and for $k=1$, we expected the black hole to have one horizon like schwarzschild black hole, but we saw that FQT gravity has the ability to play the role of electric charge and so we can have a black hole with two horizons.\\
FQT gravity has also a different behavior to TQT and Einstein gravity. It can cause a black hole with three horizons for $k=-1$ in contrast with the other two gravities.
We can also aim that the horizons of the obtained solutions for FQT theory in dS spacetimes are smaller than the ones in Einstein gravity. We know that black holes with smaller horizons can lead to more stability. So, black holes in FQT gravity are more stable than the ones in Einstein gravity.\\
We also checked thermal stability of the obtained solutions. We deduced that thermal stability are just for AdS solutions not for dS and flat ones. For different values of parameter $q$, AdS solutions show stability for $k=-1,0$ when $T_{+}>0$. we also concluded that the solutions with smaller $q$ have a larger region for thermal stability with respect to larger $q$. we also distinguished that thermal stability is independent to the value of $c$.
Quasitopological gravity could lead to interesting solutions for Reissner-Nordstr\"om black holes. It may also be attractive to study the effect of this gravity on the solutions with nonlinear electrodynamics.

\acknowledgments{We would like to thank Payame Noor University and Jahrom
University.}

\end{document}